\documentclass[pmlr]{jmlr}

\RequirePackage{graphicx}
 \usepackage{booktabs}
\usepackage{longtable}
 \usepackage{multicol}
 \usepackage{multirow}
\usepackage{tcolorbox}

\newcommand{\INSIGHT}[2]{\begin{tcolorbox}[colback=blue!5, colframe=blue!40, sharp corners, boxrule=0.5mm]\textbf{#1} #2\end{tcolorbox}}
\usepackage[dvipsnames]{xcolor}

\makeatletter
\def\set@curr@file#1{\def\@curr@file{#1}} 
\makeatother
\usepackage[load-configurations=version-1]{siunitx} 

\theorembodyfont{\upshape}
\theoremheaderfont{\scshape}
\theorempostheader{:}
\theoremsep{\newline}

\jmlrvolume{340}
\jmlryear{2026}
\jmlrworkshop{Machine Learning for Healthcare}

\title[ELF: A Family of Encoder-Free ECG-Language Models]{ELF: A Family of Encoder-Free ECG-Language Models}

\author{%
\Name{William Jongwon Han$^*$} \Email{wjhan@andrew.cmu.edu}\\
\addr Carnegie Mellon University, USA
\AND
\Name{Tony Chen$^*$}
\Email{tonyche2@andrew.cmu.edu}\\
\addr Carnegie Mellon University, USA
\AND
\Name{Chaojing Duan} \Email{chaojing.duan@ahn.org}\\
\addr Allegheny Health Network, USA
\AND
\Name{Xiaoyu Song} \Email{xiaoyuso@andrew.cmu.edu}\\
\addr Carnegie Mellon University, USA
\AND
\Name{Yihang Yao} \Email{yihangya@andrew.cmu.edu}\\
\addr Carnegie Mellon University, USA
\AND
\Name{Yuzhe Yang} \Email{yuzhey@ucla.edu}\\
\addr University of California, Los Angeles, USA
\AND
\Name{Michael A. Rosenberg} \Email{michael.a.rosenberg@cuanschutz.edu}\\
\addr University of Colorado, USA
\AND
\Name{Emerson Liu} \Email{emerson.liu@ahn.org}\\
\addr Allegheny Health Network, USA
\AND
\Name{Ding Zhao} \Email{dingzhao@andrew.cmu.edu}\\
\addr Carnegie Mellon University, USA
}

\begin{document}

\maketitle
\footnotetext[1]{$^*$ Denotes equal contribution.}

\begin{abstract}
ECG–Language Models (ELMs) extend recent advances in Multimodal Large Language Models (MLLMs) to automated ECG interpretation. However, most existing ELMs inherit Vision–Language Model (VLM) design choices and rely on pretrained ECG encoders, introducing substantial architectural and training complexity. Inspired by encoder-free VLMs, we introduce \textbf{ELF}, a family of three encoder-free ELMs that remain competitive with, and often outperform, prior state-of-the-art ELMs across two datasets despite substantially simpler architectures and training pipelines. All code and data are available at \href{https://github.com/ELM-Research/ECG-Language-Models}{\texttt{github.com/ELM-Research/ECG-Language-Models}}.
\end{abstract}

\section{Introduction}

The global volume of over 300 million ECGs recorded annually \citep{zhu_2020_automatic}, combined with the expertise required for accurate interpretation and the growing shortage of physicians \citep{aamc_2024_the}, has created demand for automated ECG analysis. Recently, the automation of ECG interpretation has advanced from classification-based models \citep{rajpurkar2017cardiologistlevelarrhythmiadetectionconvolutional} to ECG–Language Models (ELMs) \citep{zhao2024ecgchatlargeecglanguagemodel}. 

\begin{figure}[t]
    \centering
    \includegraphics[width=1\linewidth]{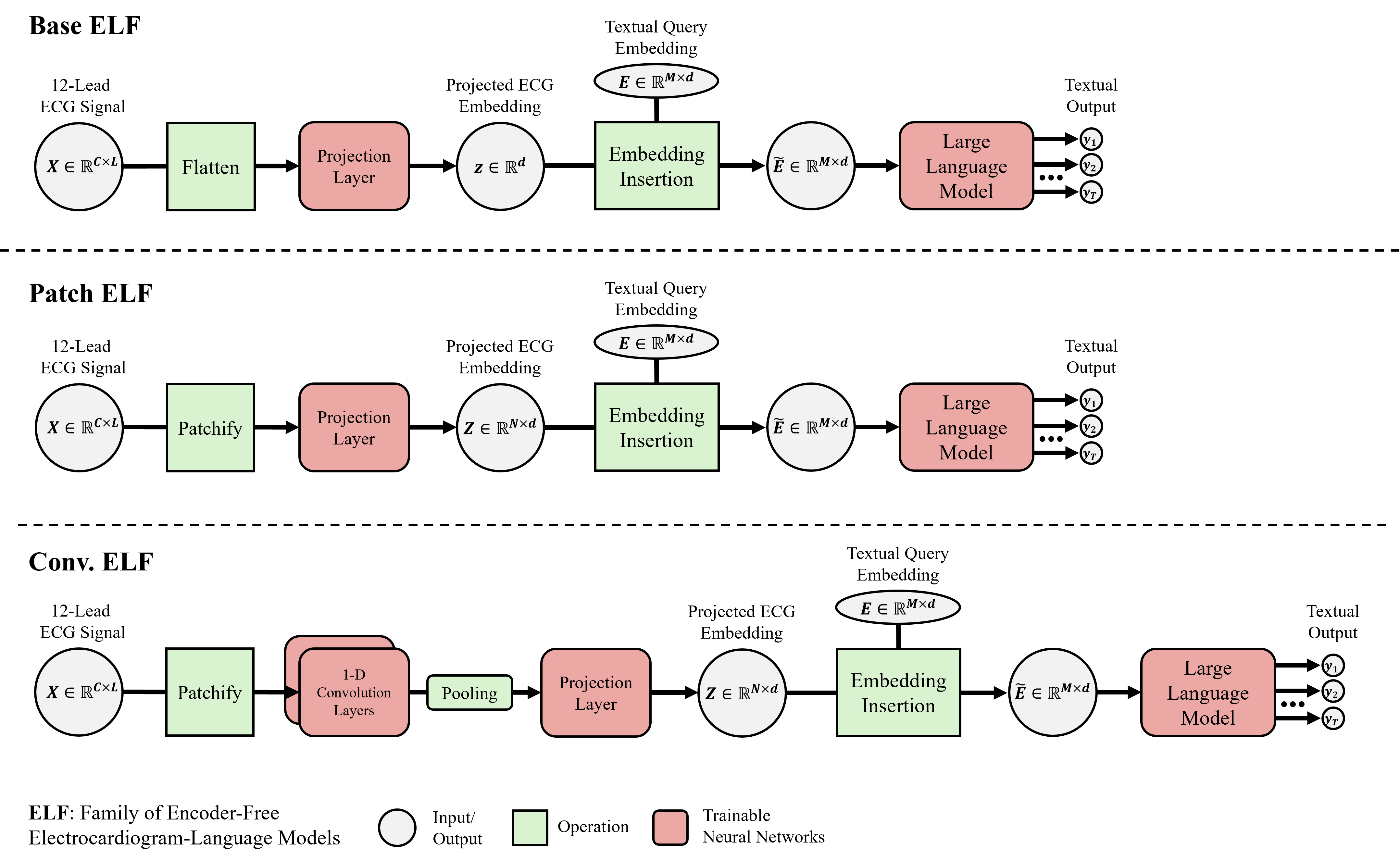}
    \caption{\textbf{The Family of ELF}. We present three encoder-free ELM architectures: Base ELF, Patch ELF, and Conv.\ ELF. Base ELF maps the full ECG into a single embedding token, Patch ELF represents the ECG as a sequence of non-overlapping patch tokens, and Conv.\ ELF augments patch-based tokenization with lightweight temporal convolutions before projection into the LLM hidden space.}
    \label{fig:ef_main}
\end{figure}

In Vision-Language Models (VLMs), RGB images are typically processed by a large neural network (vision encoder), which is first pretrained on internet-scale image datasets with its own learning objective \citep{bai2025qwen25vltechnicalreport}. The pretrained encoder is then frozen and paired with a projection layer that maps its output to a latent representation compatible with the Large Language Model (LLM) \citep{liu2023visualinstructiontuning}. Similarly, many ELM architectures include an ECG encoder that is separately pretrained on large collections of ECG data \citep{li2025anyecgchatgeneralistecgmllmflexible}. In practice, developing an effective encoder, whether for ECG or vision data, is a complex process that requires careful feature engineering and tuning of the architecture, learning objectives, and training procedures.

To address these complexities in the VLM domain, an encoder-free VLM called Fuyu-8B \citep{bavishi_2023_fuyu8b} was introduced. Unlike conventional VLMs, Fuyu-8B omits a dedicated vision encoder and instead adds a simple linear projection for patched images. The projection layer and decoder-only transformer are trained jointly in an end-to-end manner using an autoregressive objective. Fuyu-8B demonstrated that a vision encoder is not necessary to achieve competitive performance on VLM benchmarks compared to state-of-the-art (SOTA) VLMs at the time that relied on intricate vision encoders.

Taking inspiration from Fuyu-8B, we introduce ELF, a family of three encoder-free ELM architectures: (1) Base ELF, (2) Patch ELF, and (3) Conv. ELF. We compare ELF against four state-of-the-art ELMs that rely on more complex encoders and training pipelines: OpenTSLM \citep{langer2025opentslmtimeserieslanguagemodels}, SLIP \citep{chen2026learningtransferablesensormodels}, ST-MEM \citep{na2024guidingmaskedrepresentationlearning, han2025signalimagesymbolicexploring}, and MERL \citep{liu2024zeroshotecgclassificationmultimodal, han2025signalimagesymbolicexploring}. Across modern ECG-language benchmarks, including ECG-QA-CoT \citep{langer2025opentslmtimeserieslanguagemodels, oh2023ecgqacomprehensivequestionanswering} and ECG-Instruct 45K \citep{zhao2024ecgchatlargeecglanguagemodel}, ELF remains consistently competitive and often outperforms these baselines. To better understand the factors underlying ELF’s performance, we conduct additional experiments varying the LLM backbone, training duration, number of tokens, and trainable components. We summarize our main contributions below:\\
\textbf{1.} We introduce \textbf{Base ELF}, the simplest member of the ELF family, which directly flattens a 12-lead, 10-second ECG and projects it as a single representative ECG token into the LLM input space alongside the textual query embedding.\\
\textbf{2.} We introduce \textbf{Patch ELF}, which extends Base ELF by partitioning the ECG into non-overlapping temporal patches and projecting each patch as a separate token, allowing us to study whether patch-based tokenization improves performance.\\
\textbf{3.} Lastly, we introduce \textbf{Conv. ELF}, which builds on Patch ELF by adding lightweight temporal convolutions before projection, enabling us to examine whether simple local inductive biases provide additional benefit.

Based on these contributions, we summarize our main findings below:
\begin{tcolorbox}[colback=blue!5, colframe=blue!40, sharp corners, boxrule=0.5mm]
\textbf{Summary of Main Findings}
\begin{itemize}
    \item All ELF architectures remain highly competitive with prior ELMs, often outperforming prior baselines despite using substantially simpler architectures and training pipelines. (\S\ref{sec:baselines_ecgqa})
    \item Pretrained ECG encoders yield only modest performance gains while requiring substantially more training compute and larger models than ELF. (\S\ref{sec:baselines_ecgqa})
    \item Increasing the number of encoder tokens from $N=50$ to $N=100$ for both Patch and Conv. ELF does not improve performance, indicating no measurable benefit from a larger token budget. (\S\ref{sec:num_encoder})
    \item Selective training results show that Patch ELF depends primarily on the pretrained LLM backbone rather than the projection layer. (\S\ref{sec:selective})
    \item Patch ELF performance drops substantially under inference-time ECG perturbations, suggesting that coherent ECG inputs are important for strong performance. (\S\ref{sec:perturbations})
\end{itemize}
\end{tcolorbox}

\subsection*{Generalizable Insights about Machine Learning in the Context of Healthcare}
This study suggests that, in ECG-Language modeling, added architectural and training complexity should not be assumed to improve performance. Across our experiments, ELF remain highly competitive with substantially more complex baselines (e.g., OpenTSLM \citep{langer2025opentslmtimeserieslanguagemodels}, SLIP \citep{chen2026learningtransferablesensormodels}), while ELMs with pretrained ECG encoders (e.g., ST-MEM \citep{na2024guidingmaskedrepresentationlearning}, MERL \citep{liu2024zeroshotecgclassificationmultimodal}) yield only modest improvements despite requiring much greater compute.

\section{Related Work}
\subsection{ECG-Language Models}
The automation of ECG analysis is advancing toward generative approaches with ELMs \citep{han2025signalimagesymbolicexploring}. Previous ECG classification systems are limited to outputting probability scores for a fixed set of diagnostic labels \citep{qiu-etal-2023-transfer, hannun_2019_cardiologistlevel, pmlr-v225-qiu23a, Martin2021RealtimeFS, Strodthoff2021DeepLF, liu2024zeroshotecgclassificationmultimodal, na2024guidingmaskedrepresentationlearning, choi2023ecgbert, jin2025readingheartlearningecg}. In contrast, ELMs can process both text and ECG inputs and generate free-form textual responses conditioned on the given input. Thus, ELMs not only inherit the capabilities of classification systems but also extend them to more general tasks, such as conversational question answering and clinically relevant applications like ECG waveform analysis, patient context inference, and treatment planning. This broad potential motivates the transition from classification-based ECG analysis to generative approaches with ELMs.

Although promising, ELMs remain in their early stages, with most prior works still exploring different ECG representations, architectures, learning objectives, and training regimes \citep{lan2025gemempoweringmllmgrounded,zhao2024ecgchatlargeecglanguagemodel,song2025retrievalaugmentedgenerationelectrocardiogramlanguagemodels}. Following the architectural pattern of VLMs, many studies first train a strong ECG-specific encoder, then pair the pretrained encoder with a pretrained LLM for natural language generation (NLG) \citep{li2025anyecgchatgeneralistecgmllmflexible}. However, this approach requires substantial ECG data and computational resources to train an effective encoder. To address this limitation, recent works have proposed more efficient designs. MEIT \citep{wan2024meitmultimodalelectrocardiograminstruction}, for instance, employs a lightweight ECG encoder composed of several 1-D convolutional layers with batch normalization \citep{ioffe2015batchnormalizationacceleratingdeep}, ReLU \citep{fukushima_1969_visual}, and average pooling, trained end-to-end with the LLM. Another approach explores non-learning-based methods, such as applying Byte-Pair Encoding (BPE) \citep{Gage1994ANA} to tokenize the ECG signal and directly concatenate it with text query tokens \citep{han2024ecgbytetokenizerendtoendgenerative}. We benchmark ELF against several recent ELM methods \citep{langer2025opentslmtimeserieslanguagemodels, na2024guidingmaskedrepresentationlearning, liu2024zeroshotecgclassificationmultimodal, han2025signalimagesymbolicexploring, chen2026learningtransferablesensormodels} and show that it consistently matches or outperforms them. This suggests strong performance on current benchmarks does not necessarily require complex encoders or elaborate training pipelines.

\subsection{Encoder-Free Vision-Language Models}
\label{rl:ef-vl}
Fuyu-8B \citep{bavishi_2023_fuyu8b} is one of the first encoder-free VLMs developed. The authors argue that existing VLMs are overly complex, typically requiring separate image encoders, multi-stage training, and intricate connector modules. Fuyu-8B employs a simple decoder-only transformer where image patches are linearly projected into the first transformer layer. When benchmarked against state-of-the-art VLMs at the time, Fuyu-8B achieved competitive performance. We adopt the same motivation in developing ELF.

Since Fuyu-8B, there have been several advances in encoder-free VLMs \citep{diao2024unveilingencoderfreevisionlanguagemodels, chameleonteam2025chameleonmixedmodalearlyfusionfoundation, diao2025evev2improvedbaselinesencoderfree}. However, we note that the “connector” between the image and the LLM has grown increasingly complex, resembling a vision encoder. This trend undermines the original motivation behind encoder-free VLMs: \textit{simplicity}. To revisit this design principle in ECG-language modeling, we introduce a family of three encoder-free ELMs: Base ELF, Patch ELF, and Conv. ELF. Base ELF is the simplest variant, directly flattening a 12-lead, 10-second ECG and mapping it into the LLM embedding space with a single linear projection. Patch ELF extends this design by splitting the ECG into non-overlapping temporal patches and projecting each patch as a separate input token, enabling us to test whether patch-based tokenization improves performance. Conv. ELF further augments Patch ELF with lightweight temporal convolutions prior to projection, incorporating simple inductive biases to examine whether local feature extraction provides additional benefit. We provide a more comprehensive description of each ELF variant in Section~\ref{sec:elf}. We deliberately choose these three designs to test whether controlled increases in architectural complexity yield meaningful gains in ELM performance.

\section{Methods}
In this section, we describe the datasets, ELMs, and learning objectives used in our study. We also detail the training and evaluation procedures to support reproducibility.

\subsection{Datasets}
We use preprocessed 12-lead, 10-second ECGs sampled at 250 Hz with 70:30 training/testing splits across two datasets comprising ECGs and text: ECG-Instruct 45K \citep{zhao2024ecgchatlargeecglanguagemodel} and ECG-QA-CoT \citep{oh2023ecgqacomprehensivequestionanswering, langer2025opentslmtimeserieslanguagemodels}.
These datasets are derived from ECG recordings originating from MIMIC-IV-ECG \citep{johnson_2023_mimiciv} and PTB-XL \citep{wagner_ptb-xl_2020}. Lastly, we provide example instances from ECG-Instruct 45K and ECG-QA-CoT in Figures~\ref{fig:ecg-instruct-inst} and ~\ref{fig:ecg-qa-inst} of Appendix~\ref{app:data_inst}.

\paragraph{ECG-Instruct 45k}
ECG-Instruct 45K \citep{zhao2024ecgchatlargeecglanguagemodel} is an instruction-following conversational dataset generated with GPT-4o \citep{openai2024gpt4ocard}. It consists of dialogues between a physician and a patient covering topics such as ECG interpretation, heart rate, waveform morphology, rhythm, and cardiac axis. The dataset is derived from MIMIC-IV-ECG \citep{johnson_2023_mimiciv}. In our study, we use ECG-Instruct 45K to train and evaluate the conversational capabilities of ELMs.

\paragraph{ECG-QA-CoT} ECG-QA-CoT \citep{langer2025opentslmtimeserieslanguagemodels} extends ECG-QA \citep{oh2023ecgqacomprehensivequestionanswering} by generating Chain-of-Thought (CoT) \citep{wei2023chainofthoughtpromptingelicitsreasoning} rationales with GPT-4o \citep{openai2024gpt4ocard}, conditioned on a plotted ECG image and its corresponding clinical context and question-answer pair. The clinical context consists of non-diagnostic PTB-XL metadata \citep{wagner_ptb-xl_2020}, including patient demographics, basic recording information, signal quality notes, and additional technical observations such as extra beats or presence of pacemaker. The clinical context is also used during OpenTSLM's training and evaluation \citep{langer2025opentslmtimeserieslanguagemodels}. In our study, we adopt the same train/test splits used in both OpenTSLM \citep{langer2025opentslmtimeserieslanguagemodels} and SLIP \citep{chen2026learningtransferablesensormodels} for the ECG-QA-CoT dataset.

\subsection{ELF}
\label{sec:elf}
In this section, we describe each ELM architecture in the ELF family. A high-level visualization of each architecture is shown in Figure~\ref{fig:ef_main}.

\paragraph{Base ELF}
Given an ECG signal \(X \in \mathbb{R}^{C \times L}\), where \(C\) denotes the number of leads and \(L\) the signal length, Base ELF flattens the signal into a vector
\[
x = \mathrm{vec}(X) \in \mathbb{R}^{CL},
\]
and projects it into the LLM hidden space:
\[
z = Wx + b, \qquad W \in \mathbb{R}^{d \times CL}, \; b \in \mathbb{R}^{d}.
\]
Let \(E \in \mathbb{R}^{M \times d}\) denote the textual query embeddings, where \(M\) is the number of textual input tokens. We replace the embedding of a special placeholder token \texttt{<signal>} with \(z\), yielding the combined input embedding sequence \(\tilde{E}\). The model is then trained autoregressively to generate the response tokens \(y_{1:T}\):
\[
\mathcal{L}_{\mathrm{AR}}
= - \sum_{t=1}^{T} \log P_\theta\!\left(y_t \mid y_{<t}, \tilde{E}\right),
\]
where \(\theta\) denotes the model parameters. When computing \(\mathcal{L}_{\mathrm{AR}}\), we mask all tokens except the response and end-of-sequence tokens.

\paragraph{Patch ELF}
Patch ELF extends Base ELF by partitioning the ECG signal \(X \in \mathbb{R}^{C \times L}\) into \(N = L / L_p\) non-overlapping patches of length \(L_p\). We denote the patched signal by
\[
\mathcal{X}^{(p)} = \{X_i\}_{i=1}^{N}, \qquad X_i \in \mathbb{R}^{C \times L_p}.
\]
Each patch \(X_i\) is flattened into
\[
x_i = \mathrm{vec}(X_i) \in \mathbb{R}^{CL_p},
\]
and projected using a shared linear layer:
\[
z_i = W x_i + b, \qquad W \in \mathbb{R}^{d \times CL_p}, \; b \in \mathbb{R}^{d}.
\]
Collecting the patch embeddings gives
\[
Z = [z_1,\dots,z_N] \in \mathbb{R}^{N \times d}.
\]
We then prepare \(N\) consecutive placeholder tokens \texttt{<signal$_i$>} for \(i = 1, \ldots, N\), and replace the embeddings of the corresponding placeholder tokens in the textual query embedding sequence \(E\) with \(z_i\). Unless otherwise specified, we use \(N = 50\). Patch ELF investigates whether representing the ECG as a sequence of fixed-size patches, analogous to patching in VLMs, improves downstream performance.

\paragraph{Conv.\ ELF}
Conv.\ ELF extends Patch ELF by replacing the linear patch projection with a series of convolutional layers. Given the patched ECG signal \(\mathcal{X}^{(p)} = \{X_i\}_{i=1}^{N}\), where each patch \(X_i \in \mathbb{R}^{C \times L_p}\), Conv.\ ELF applies two 1D convolutions along the temporal dimension, treating the \(C\) leads as input channels. Let
\[
h_i = G(X_i),
\]
where \(G\) denotes a two-layer convolutional module, with a ReLU activation \citep{fukushima_1969_visual} applied after each convolution. The resulting feature map is globally average pooled and projected into the LLM hidden space:
\[
z_i = W \, \mathrm{Pool}(h_i) + b, \qquad W \in \mathbb{R}^{d \times D_c}, \; b \in \mathbb{R}^{d},
\]
where \(D_c\) denotes the channel dimension after convolution and pooling. Collecting the patch embeddings gives $Z = [z_1,\dots,z_N] \in \mathbb{R}^{N \times d}$, where $N=50$.

The patch embeddings are then inserted into the textual query embedding sequence \(E\) following the same procedure as in Patch ELF. Conv.\ ELF investigates whether combining patching with lightweight temporal convolutions, and thus introducing simple inductive biases, improves ECG feature extraction and downstream performance.

\subsection{Baselines}
We provide descriptions of the baselines we compare against ELF. We utilize four strong baselines in this study: OpenTSLM \citep{langer2025opentslmtimeserieslanguagemodels}, SLIP \citep{chen2026learningtransferablesensormodels}, ST-MEM \citep{na2024guidingmaskedrepresentationlearning}, and MERL \citep{liu2024zeroshotecgclassificationmultimodal}.

\paragraph{OpenTSLM}
OpenTSLM \citep{langer2025opentslmtimeserieslanguagemodels} is a general, time-series language model (TSLM) with two variants: OpenTSLM SoftPrompt (OpenTSLM SP) and OpenTSLM Flamingo (OpenTSLM FL).
OpenTSLM SP converts the signal into soft prompt tokens interleaved throughout the input sequence, whereas OpenTSLM FL encodes the signal separately and integrates it through gated cross-attention. Both variants are trained in two stages with a mixture of data: the first learns simple question-answering from paired time-series input, and the second introduces CoT rationales, which includes the ECG-QA-CoT dataset.

\paragraph{SLIP}
SLIP \citep{chen2026learningtransferablesensormodels} is a general, sensor-language model designed to learn transferable, language-aligned representations from multivariate time-series data. It extends the CoCa framework \citep{yu2022cocacontrastivecaptionersimagetext} to the sensor domain with a Transformer-based sensor encoder, attention pooling module, text encoder, and multimodal decoder. The model is pretrained with a CLIP-style contrastive loss and an autoregressive captioning loss on over 600K paired sensor-text examples from diverse domains, then finetuned for question answering on ECG-QA-CoT using the same train/test splits as OpenTSLM \citep{langer2025opentslmtimeserieslanguagemodels}, denoted as $\text{SLIP}_{\text{SFT}}$ \citep{chen2026learningtransferablesensormodels}. In our study, we exclusively compare with $\text{SLIP}_{\text{SFT}}$ and treat it as a strong finetuned sensor baseline.

\paragraph{ST-MEM and MERL as ELMs}
We convert both ST-MEM \citep{na2024guidingmaskedrepresentationlearning} and MERL \citep{liu2024zeroshotecgclassificationmultimodal} into ELMs through a LLaVA-like architecture \citep{liu2023visualinstructiontuning}. ST-MEM is a masked autoencoder (MAE) \citep{he2021maskedautoencodersscalablevision} built on a Vision Transformer (ViT) backbone \citep{dosovitskiy2021image}, while MERL is a multimodal ECG-text representation model trained to align ECGs and reports through contrastive learning. For MERL, we use its strongest backbone, based on the ResNet101 architecture \citep{he2015deepresiduallearningimage}. The output of each encoder is projected into the LLM embedding space through a linear layer and inserted at the placeholder tokens \texttt{<signal$_i$>} for \(i = 1, \ldots, N\). The resulting model is then trained with the same autoregressive objective $\mathcal{L}_{\text{AR}}$.
During this stage, the encoder is frozen, and only the projection layer and LLM backbone are trained. For both models, a 1D adaptive average pooling layer controls the number of encoder tokens $N$, where $N=50$ unless otherwise noted. We provide a detailed table of the hyperparameters for training ST-MEM and MERL encoders in Table~\ref{tab:train_hparams_combined} of Appendix~\ref{app:hyp}. Lastly, we refer to these resulting ELMs as ST-MEM and MERL throughout the rest of the study.

\subsection{Large Language Model Backbones}
For ELF, we experiment with the Llama-3.2-1B-Instruct\footnote{\texttt{meta-llama/Llama-3.2-1B-Instruct}}, Qwen2.5-1.5B-Instruct\footnote{\texttt{Qwen/Qwen2.5-1.5B-Instruct}}, and gemma-2-2b-it\footnote{\texttt{google/gemma-2-2b-it}} checkpoints \citep{grattafiori2024llama3herdmodels, qwen2025qwen25technicalreport, gemmateam2024gemma2improvingopen} accessed via the HuggingFace API \citep{wolf2020huggingfaces}. ST-MEM and MERL ELM baselines use only the Llama-3.2-1B-Instruct LLM backbone. All LLMs are initialized with their default hyperparameters.

We use the Muon \citep{liu2025muonscalablellmtraining} optimizer with learning rate 1e-3, weight decay 5e-2, \(\beta_{1}=0.9\), \(\beta_{2}=0.95\), and \(\epsilon=1e-8\). We employ LoRA \citep{hu2021loralowrankadaptationlarge} with rank 16, \(\alpha_{LoRA}=32\), and dropout 0.05. We use a batch size of 4 with 10 second 12-lead ECGs, max input sequence length of 2048, and warm-up for 10\% of total steps across 10 epochs. We provide a detailed table of all hyperparameters used during ELM training in Table~\ref{tab:train_hparams_combined} of Appendix~\ref{app:hyp}.

\subsection{Evaluation}
We evaluate all models over three random seeds, reporting performance on the following text generation metrics: BLEU-4 \citep{Papineni2002BleuAM}, METEOR \citep{Banerjee2005METEORAA}, ROUGE-L \citep{Lin2004ROUGEAP}, and BERTScore F1 \citep{Zhang2020BERTScoreET}. We also report F1 and accuracy. We define accuracy as the proportion of generated responses that exactly match the ground truth (i.e., exact string matching). We define F1 using token-level overlap after lowercasing, removing punctuation, and normalizing whitespace. For each ground truth–prediction pair, we compute precision and recall from the overlapping tokens and take their harmonic mean. The final F1 score is the average of these per-sample token-level F1 values across the dataset. All metrics are in percentage form.

\section{Results}
In this section, we present results on the ECG-QA-CoT and ECG-Instruct 45K datasets and report several ablation studies.
\newpage
\begin{figure}
    \centering
    \includegraphics[width=1\linewidth]{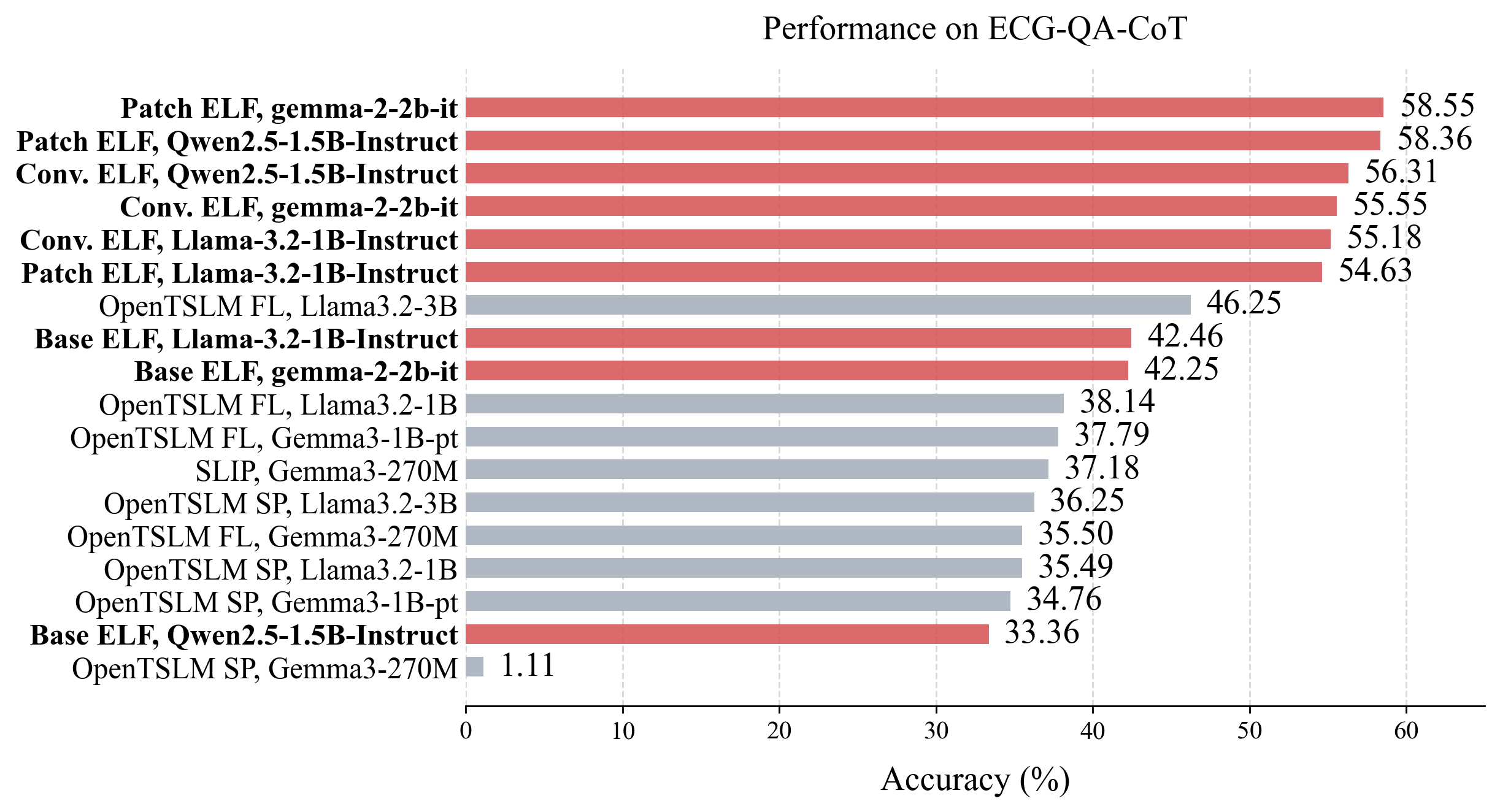}
    \caption{Comparing performance on ECG-QA-CoT between various baselines (e.g., OpenTSLM \citep{langer2025opentslmtimeserieslanguagemodels} and SLIP \citep{chen2026learningtransferablesensormodels}) and ELF. ELMs from the ELF family are colored in \textcolor{red}{red}. A tabular summary of these results, including additional metrics and standard deviations, is provided in Table~\ref{tab:ecgqa_cot_opentslm} of Appendix~\ref{app:add_results}.}
    \label{fig:main}
\end{figure}

\subsection{Baselines on ECG-QA-CoT and ECG-Instruct 45K}
\label{sec:baselines_ecgqa}
\paragraph{ECG-QA-CoT} We first compare the accuracies of all methods on ECG-QA-CoT in Figure~\ref{fig:main}. While the figure highlights accuracy only, the full results, including BLEU-4, ROUGE-L, METEOR, and BERTScore F1, are provided in Table~\ref{tab:ecgqa_cot_opentslm} in Appendix~\ref{app:add_results}. Overall, the ELF family is highly competitive, with Base, Patch, and Conv. ELF occupying eight of the top nine positions against prior ELM baselines such as OpenTSLM \citep{langer2025opentslmtimeserieslanguagemodels} and SLIP \citep{chen2026learningtransferablesensormodels}. Patch ELF with gemma-2-2b-it is the strongest model, achieving 58.55\% accuracy. It substantially outperforms the best non-ELF baseline, OpenTSLM FL with Llama3.2-3B, despite the latter using a more complex architecture and training strategy (i.e., curriculum learning), and a larger pretrained backbone. Patch ELF with the Qwen2.5-1.5B-Instruct backbone follows closely at 58.36\%. In contrast, Base ELF with Qwen2.5-1.5B-Instruct performs worst among the ELF variants at 33.36\% accuracy. Notably, despite being the simplest variant of ELF, Base ELF with the Llama-3.2-1B-Instruct and gemma-2-2b-it backbones still outperforms SLIP and all OpenTSLM variants except OpenTSLM FL with Llama3.2-3B. Nonetheless, Patch ELF and Conv.\ ELF consistently outperform Base ELF, suggesting that patching and lightweight convolutional layers are beneficial on ECG-QA-CoT. Taken together, these results show that simple encoder-free designs can remain highly competitive with, and often outperform, more complex prior ELM pipelines.

\INSIGHT{Findings 1}{ELF occupies 8 of the top 9 positions on ECG-QA-CoT despite having simpler architectures and training procedures than baselines.}

\begin{figure}
    \centering
    \begin{minipage}[t]{0.49\textwidth}
        \centering
        \includegraphics[width=\linewidth]{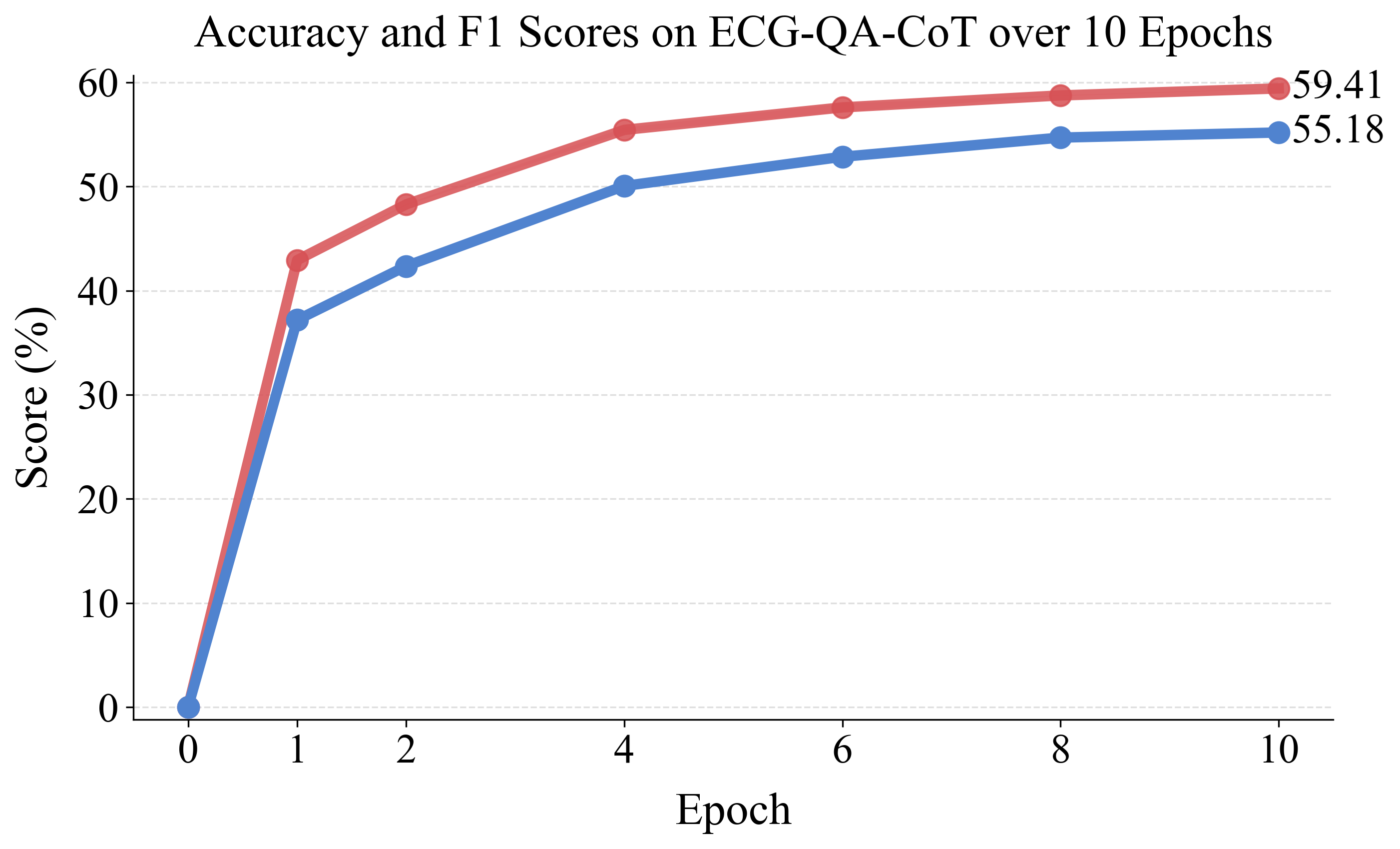}
    \end{minipage}
    \hfill
    \begin{minipage}[t]{0.49\textwidth}
        \centering
        \includegraphics[width=\linewidth]{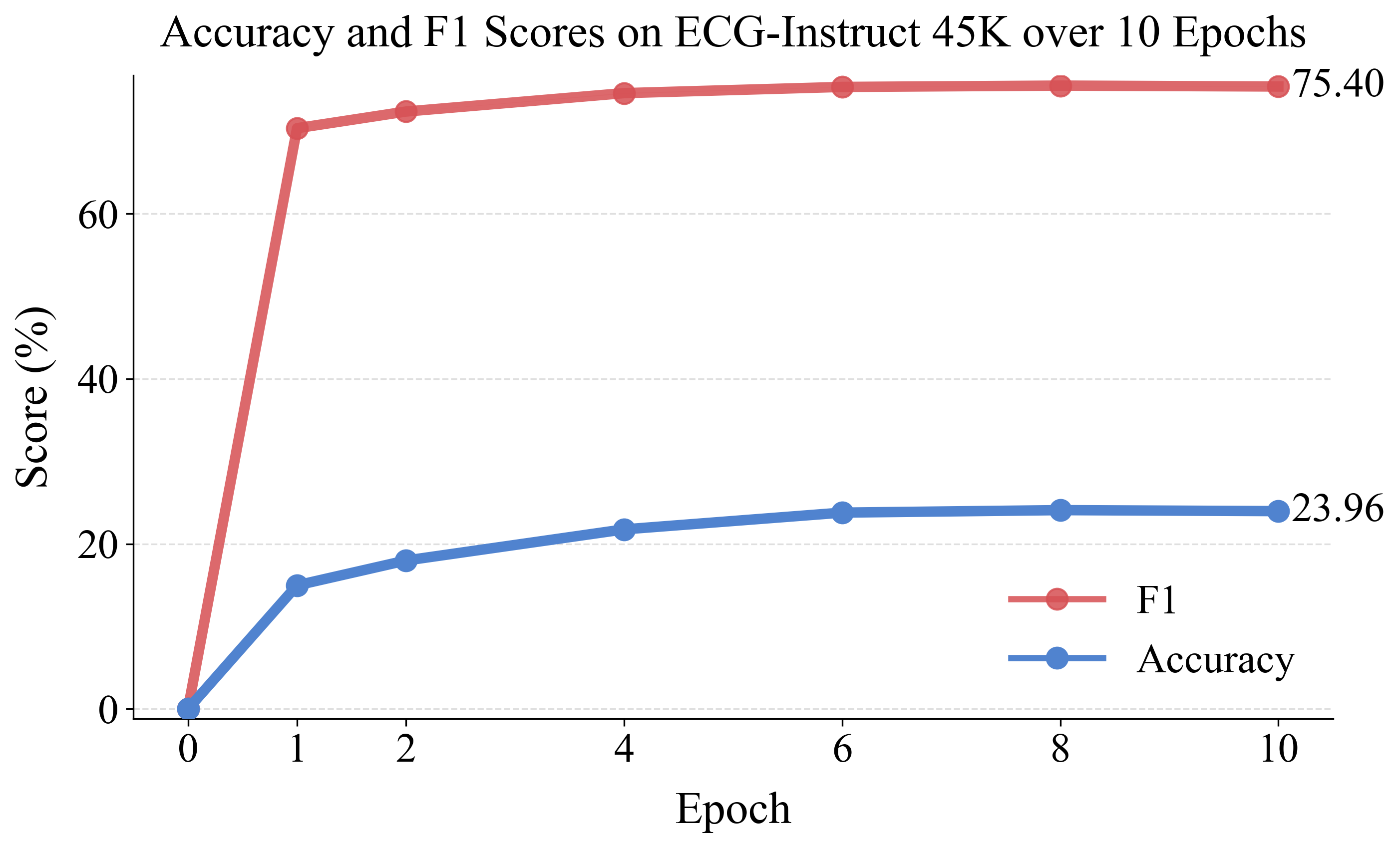}
    \end{minipage}
    \caption{
        We plot the accuracy (\textcolor{blue}{Blue}) and F1 (\textcolor{red}{Red}) scores on the test set of ECG-QA-CoT (\textbf{Left}) and ECG-Instruct 45K (\textbf{Right}) over 10 training epochs for Conv. ELF.}
    \label{fig:analysis}
\end{figure}

\paragraph{ECG-Instruct 45K} Figure~\ref{fig:analysis2} and Table~\ref{tab:ecg_instruct} (see Appendix~\ref{app:add_results}) compare ELF variants against ST-MEM and MERL on ECG-Instruct 45K. All ELM use the Llama-3.2-1B-Instruct LLM backbone. MERL achieves the highest accuracy (26.53\%), with ST-MEM close behind. Conv.\ ELF narrows this gap considerably, reaching 23.96\% accuracy, while Base ELF and Patch ELF trail at roughly 20\% accuracy. Interestingly, while Base ELF consistently underperforms Patch ELF and Conv.\ ELF on ECG-QA-CoT, it outperforms Patch ELF on ECG-Instruct 45K, suggesting that the effectiveness of ELM design choices is dataset- and task-dependent. We note that accuracy is uniformly low on this dataset because ECG-Instruct 45K is conversational and substantially more free-form than ECG-QA-CoT. We therefore emphasize the F1 score in Figure~\ref{fig:analysis}, together with the additional metrics (i.e., BLEU-4, ROUGE-L, METEOR, BERTScore F1) reported in Table~\ref{tab:ecg_instruct}, as a more complete view of model performance.

\begin{figure}
    \centering
    \begin{minipage}[t]{0.49\textwidth}
        \centering
        \includegraphics[width=\linewidth]{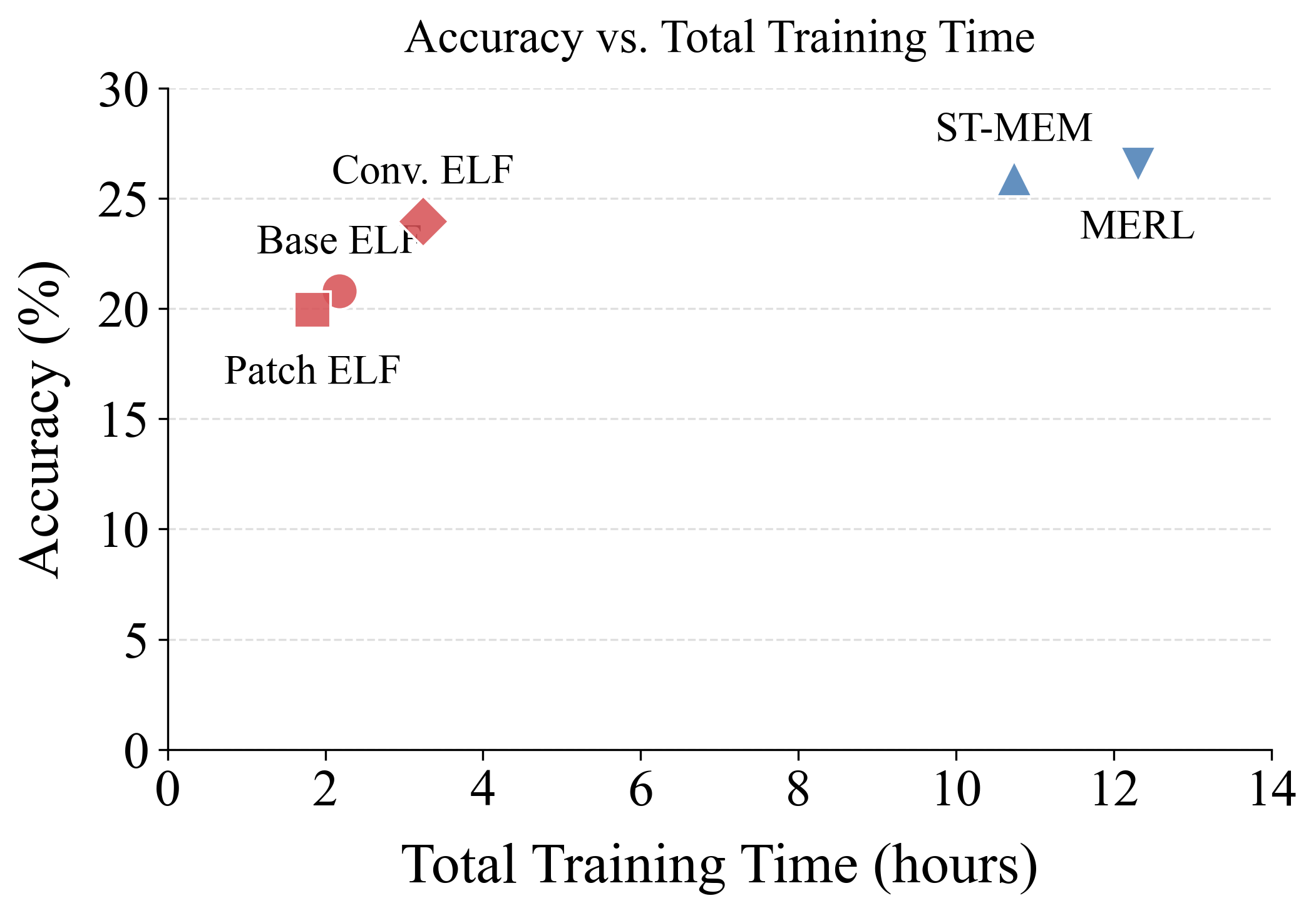}
    \end{minipage}
    \hfill
    \begin{minipage}[t]{0.49\textwidth}
        \centering
        \includegraphics[width=\linewidth]{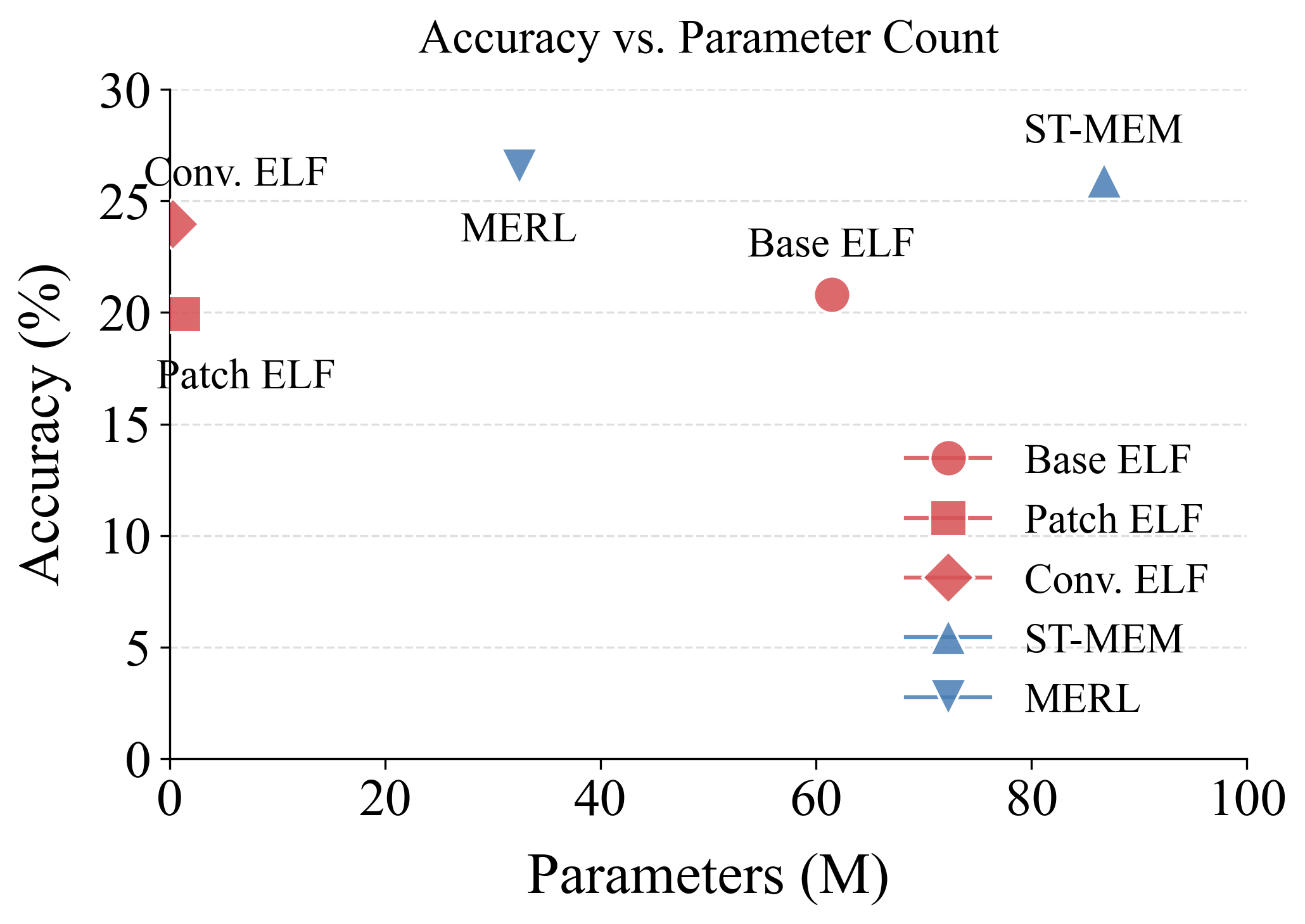}
    \end{minipage}
    \caption{
        Accuracy vs.\ total training time (\textbf{Left}) and accuracy vs.\ parameter count (\textbf{Right}) for ELF variants (\textcolor{red}{Red}) and baseline ELMs (\textcolor{blue}{Blue}) on ECG-Instruct 45K. Parameter counts only include non-LLM components.}
    \label{fig:analysis2}
\end{figure}

\paragraph{Effect of Training Epochs}
\label{sec:epochs}
We evaluate Conv.\ ELF with the Llama-3.2-1B-Instruct LLM backbone across selected epochs on the test sets of ECG-QA-CoT and ECG-Instruct 45K, shown in the left and right plots of Figure~\ref{fig:analysis}, respectively. On both datasets, performance improves rapidly in the early epochs before beginning to plateau, indicating that Conv.\ ELF should be trained close to convergence for a fair assessment. On ECG-QA-CoT, both F1 and accuracy increase from epoch 1 to epoch 10, starting at 37.16\% accuracy after epoch 1 to 55.18\% accuracy at epoch 10. On ECG-Instruct 45K, performance also improves early, with smaller changes after epoch 4. The highest scores are observed at epoch 8, with 75.51\% F1 and 24.08\% accuracy, followed by a slight decrease at epoch 10.

\paragraph{Training Efficiency} Figure~\ref{fig:analysis2} reveals that this performance gap comes at a steep efficiency cost. In terms of total training time, all three ELF variants train in under 4 hours, whereas ST-MEM and MERL require 10-12 hours. This 3-7$\times$ increase in total training time is attributed largely to the time it takes to pretrain the ST-MEM \citep{na2024guidingmaskedrepresentationlearning} and MERL \citep{liu2024zeroshotecgclassificationmultimodal} encoders. To view the total training times broken down for each ELM, please see Table~\ref{tab:training_time} in the Appendix. In terms of parameter count, the Patch and Conv. ELF convolution and projection layers contain fewer than 1M trainable parameters, compared to roughly 30M for MERL and 85M for ST-MEM. Conv.\ ELF remains well below 5M parameters while recovering the majority of the accuracy gap with the pretrained encoders. These results suggest that pretrained ECG encoders offer diminishing returns relative to their computational overhead, and that lightweight projections can achieve competitive performance at a fraction of the compute.

\INSIGHT{Findings 2}{On ECG-Instruct 45K, ELMs with pretrained ECG encoders yield only a modest accuracy gain over ELF while incurring 3-7$\times$ higher training cost and generally larger model footprints.}

\subsection{Ablation Study}
\begin{figure}
    \centering
    \begin{minipage}[t]{0.45\textwidth}
        \centering
        \includegraphics[width=\linewidth]{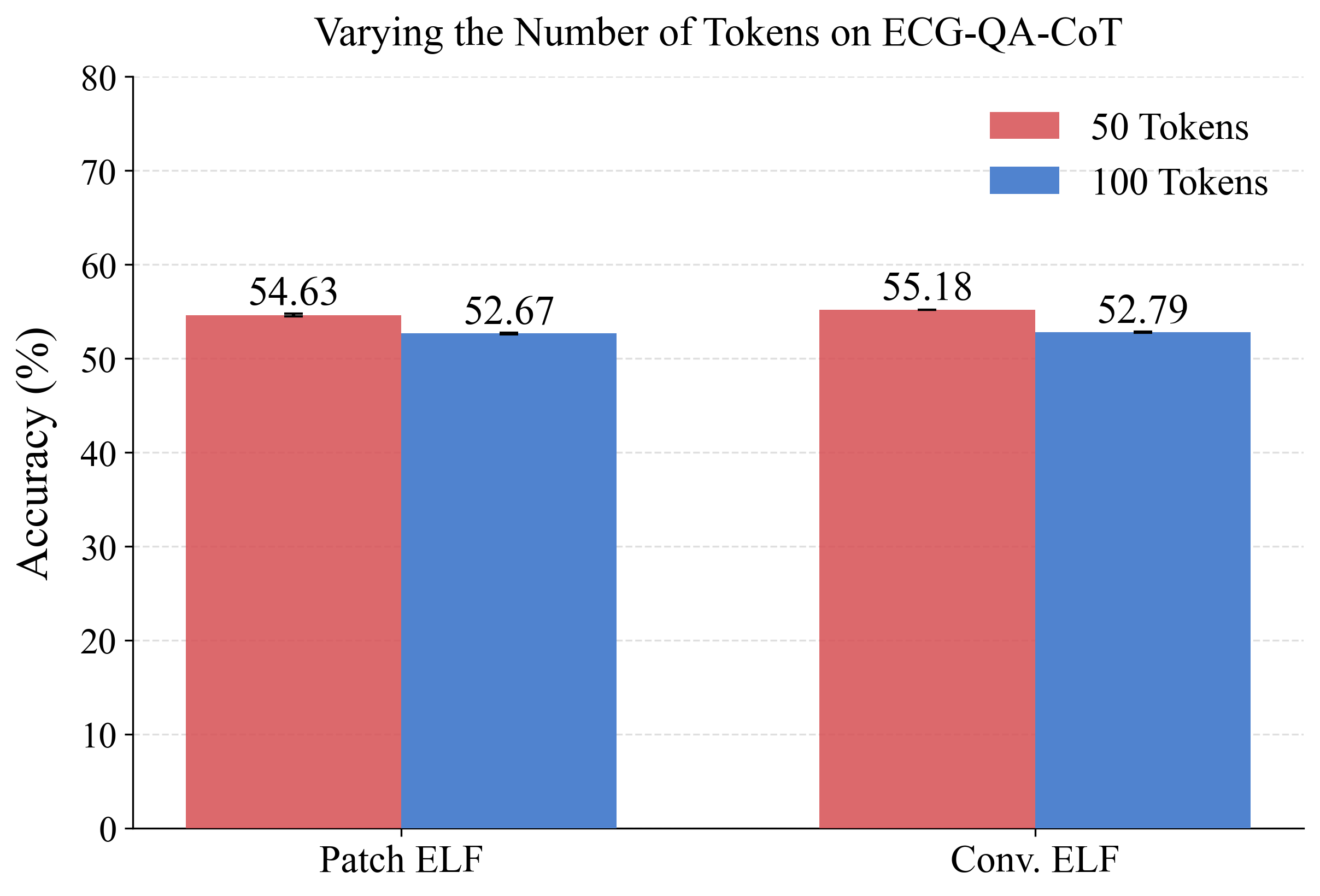}
    \end{minipage}
    \hfill
    \begin{minipage}[t]{0.53\textwidth}
        \centering
        \includegraphics[width=\linewidth]{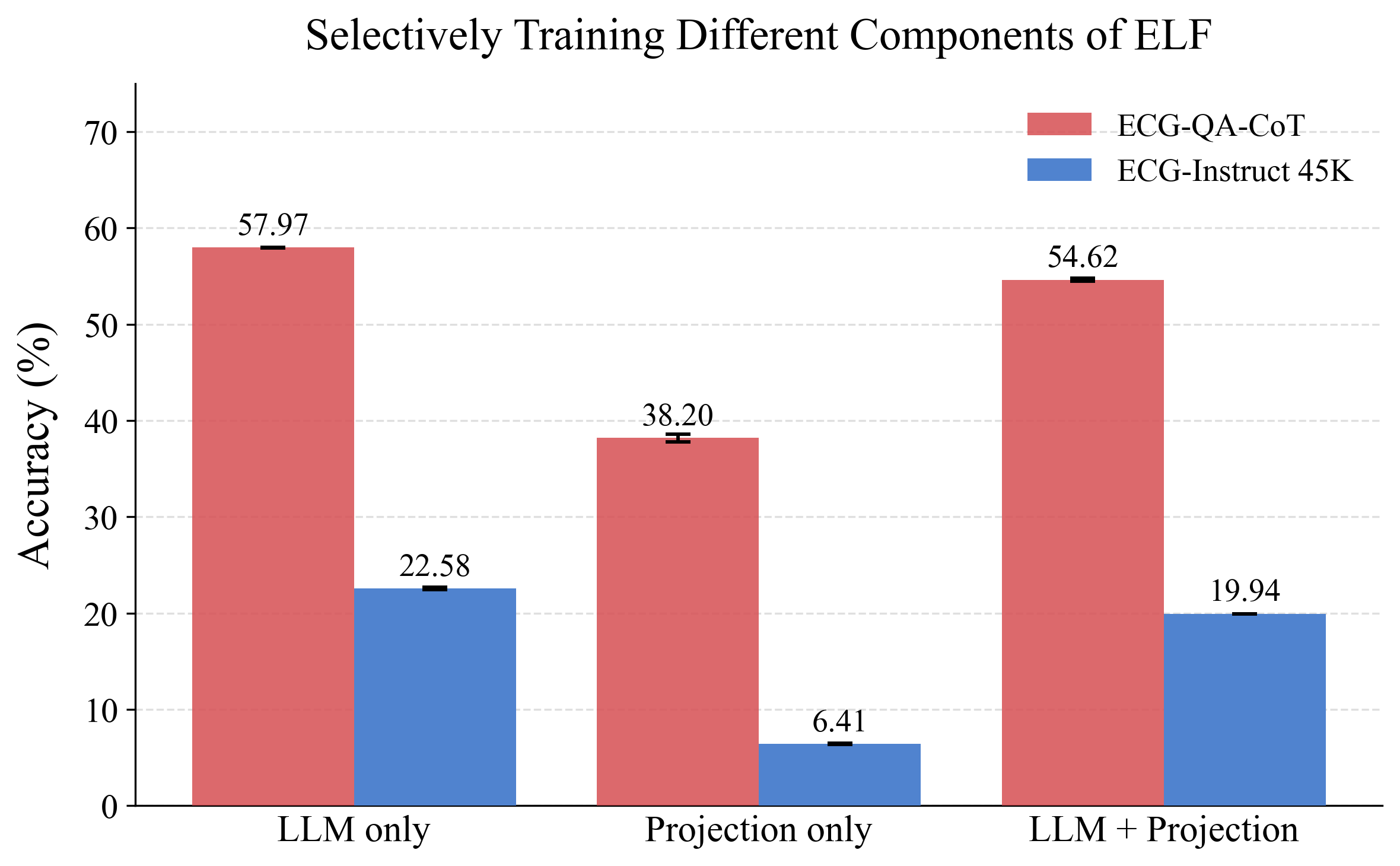}
    \end{minipage}
    \caption{(\textbf{Left}) Observing the performances on ECG-QA-CoT when varying the number of tokens $N$ for Patch and Conv. ELF. (\textbf{Right}) Selectively training Patch ELF's components on ECG-QA-CoT and ECG-Instruct 45K. LLM only, Projection only, and LLM + Projection indicate that the given component was trained while all others were frozen.
        }
    \label{fig:analysis3}
\end{figure}

\paragraph{Number of Tokens $N$}
\label{sec:num_encoder}
In the left plot of Figure~\ref{fig:analysis3}, we train and evaluate Patch ELF and Conv.\ ELF with the Llama-3.2-1B-Instruct backbone on ECG-QA-CoT using $N=100$, and compare accuracy to the default $N=50$ setting. Increasing the number of tokens does not improve performance, despite offering greater representational capacity. For additional metrics, please view Table~\ref{tab:patch_conv_elf} in the Appendix.

\INSIGHT{Findings 3}{On ECG-QA-CoT, increasing the number of encoder tokens from $N=50$ to $N=100$ does not improve performance for either Patch ELF or Conv.\ ELF. In this setting, a larger token budget provides no measurable benefit over $N=50$.}

\paragraph{Selectively Training Patch ELF Components}
\label{sec:selective}
We study the effect of selectively training Patch ELF components on the ECG-QA-CoT and ECG-Instruct 45K datasets in the right plot of Figure~\ref{fig:analysis3}. LLM only, Projection only, and LLM + Projection indicate that the given component was trained while all others were frozen. We use the Llama-3.2-1B-Instruct LLM backbone. Training only the LLM backbone while freezing the projection layer achieves the highest accuracy and consistently outperforms across most metrics (see Tables~\ref{tab:freeze_ecgqa} and \ref{tab:freeze_ecginstruct} in Appendix). Conversely, training only the projection layer while freezing the LLM yields the lowest accuracy. Training both components together slightly underperforms the LLM-only setting. We also repeat this experiment with the Gemma-2-2B-it backbone and observe similar trends, as shown in Table~\ref{tab:freeze_ecgqa}. These results suggest downstream performance in ELMs on both the ECG-QA-CoT and ECG-Instruct 45K datasets is predominantly driven by the LLM backbone.

\INSIGHT{Findings 4}{Selective training results on ECG-QA-CoT and ECG-Instruct 45K show that Patch ELF performance is driven primarily by the pretrained LLM backbone, with the LLM-only setting achieving the strongest performance across two LLM backbones.}

\subsection{ELF's Performance Under Perturbations}
\label{sec:perturbations}
\begin{figure}
    \centering
    \includegraphics[width=0.75\linewidth]{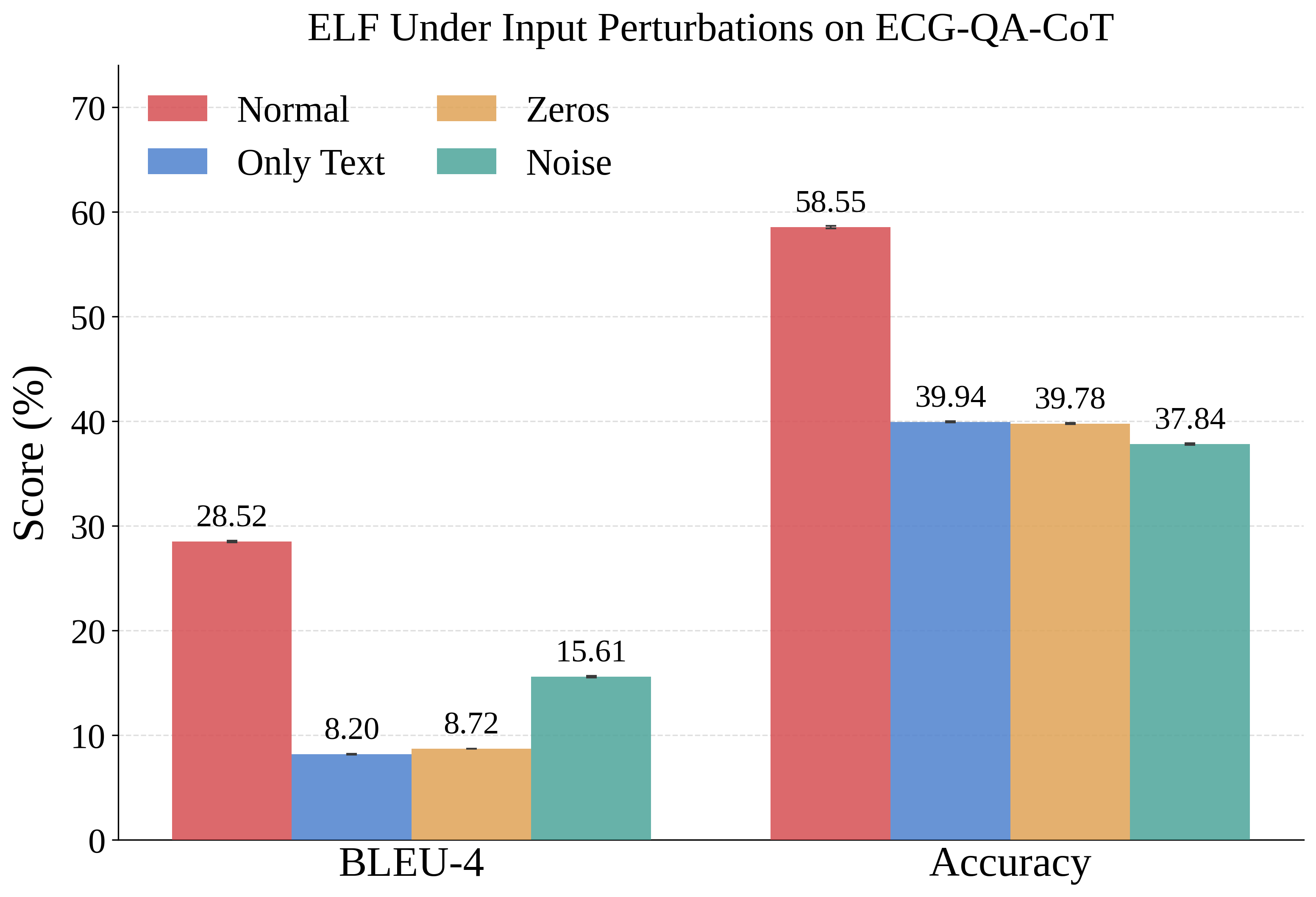}
    \caption{Patch ELF performance under three input perturbation settings on the ECG-QA-CoT dataset, using Gemma-2-2B-it as the LLM backbone.}
    \label{fig:perturb}
\end{figure}

We evaluate Patch ELF with the gemma-2-2b-it backbone under three inference-time perturbations on the ECG-QA-CoT dataset in Figure~\ref{fig:perturb}. The model is trained on normal ECG signals from ECG-QA-CoT, with perturbations applied only during inference. The three conditions are defined as follows:\\
\textbf{Noise}: A tensor of the same shape as the regular ECG signal filled with values sampled from a Gaussian distribution is used as input, alongside the textual query.\\
\textbf{Zeros Tensor}: A tensor of the same shape as the regular ECG signal filled with zeros is used as input, alongside the textual query.\\
\textbf{Only Text}: The ECG signal is completely omitted and only the textual query is used as input.\\
We note only the input is modified for each perturbation, and the architecture of ELF remains the same.

As shown in Figure~\ref{fig:perturb}, all perturbations degrade the performance of Patch ELF with the gemma-2-2b-it backbone on ECG-QA-CoT, with the largest drop observed for BLEU-4. While this suggests that ELF requires coherent ECG signal inputs to perform well on ECG-QA-CoT, these perturbation results alone do not establish ECG signal understanding and motivate further analyses of how ELF uses ECG information. Additional metrics are reported in Table~\ref{tab:perturbation_results} in the Appendix.

\INSIGHT{Findings 5}{Patch ELF performance drops substantially under inference-time ECG perturbations on ECG-QA-CoT, suggesting that coherent ECG inputs are important for strong performance.}

\section{Discussion}
As automated ECG analysis increasingly shifts from classification-based approaches to generative ones with ELMs, we introduce ELF, a family of three encoder-free ELM architectures in which each variant builds naturally on the previous one. We first introduce Base ELF, the simplest variant, which directly flattens the ECG signal and projects it with a linear layer into the LLM input space alongside the textual query. Despite its simplicity, Base ELF outperforms strong baselines such as OpenTSLM \citep{langer2025opentslmtimeserieslanguagemodels} and SLIP \citep{chen2026learningtransferablesensormodels} on ECG-QA-CoT. We then introduce Patch ELF and Conv.\ ELF. Patch ELF extends Base ELF by partitioning the ECG signal into non-overlapping patches and projecting each patch separately, while Conv.\ ELF further extends Patch ELF by adding temporal 1D convolutions before projection. Patch ELF and Conv.\ ELF also outperform all strong baselines on ECG-QA-CoT. Overall, ELF occupies eight of the top nine positions on this dataset despite using substantially simpler architectures and training procedures than prior baselines.

We also train and evaluate ELF on ECG-Instruct 45K alongside strong ELM baselines that use pretrained ECG encoders, namely ST-MEM \citep{na2024guidingmaskedrepresentationlearning} and MERL \citep{liu2024zeroshotecgclassificationmultimodal}. Although ST-MEM and MERL achieve higher performance than ELF on this dataset, they do so at a substantial computational cost, requiring 3--5$\times$ more training compute due to extensive pretraining on MIMIC-IV-ECG. In contrast, ELF requires no ECG encoder pretraining and only simple projection or lightweight convolutional layers, making it an attractive option in resource-constrained settings.

Across both datasets, we also find that added complexity within the ELF family does not consistently translate to better performance. On ECG-QA-CoT, Patch ELF, despite being architecturally simpler than Conv.\ ELF, often performs better. On ECG-Instruct 45K, Base ELF is able to outperform Patch ELF. Taken together, these results, along with ELF outperforming many strong baselines such as OpenTSLM and SLIP, suggest that simple encoder-free designs may already be sufficient to perform well on current ECG-language benchmarks.

We further examine the role of complexity by increasing the number of tokens $N$ in Patch ELF and Conv.\ ELF from 50 to 100. On ECG-QA-CoT, this leads to a slight degradation in performance, indicating that a larger token budget does not necessarily improve results.

Lastly, we investigate what may be driving ELF’s strong performance. By selectively training and freezing different components of Patch ELF on ECG-QA-CoT, we find that training only the LLM backbone while freezing the projection layer yields better performance than jointly training both the projection layer and the LLM, which is our default setting. This suggests that performance on ECG-QA-CoT may depend largely on the pretrained LLM backbone. We leave this direction for further study.

Overall, we hope this work contributes to the ELM community by introducing three new encoder-free ELM architectures and by providing further insight into what is, and may not be, necessary for strong performance on current ECG-language benchmarks.

\paragraph{Limitations}
Our study has several limitations. First, we evaluate ELF on two datasets, ECG-Instruct 45K and ECG-QA-CoT, which are derived from MIMIC-IV-ECG and PTB-XL, respectively. Evaluating on additional ECG-language datasets from diverse clinical populations would strengthen the generalizability of our findings. Second, all experiments use 12-lead, 10-second ECGs sampled at 250 Hz. Performance on recordings with different lead configurations, durations, or sampling rates is not examined. Third, we rely on standard text generation metrics (BLEU-4, ROUGE-L, METEOR, BERTScore F1) alongside token-level F1 and exact-match accuracy. These metrics measure surface-level textual similarity and do not directly assess clinical correctness. Lastly, we do not conduct any clinical validation or evaluate ELF in a real-world deployment setting.


\bibliography{sample}

\newpage
\appendix
\section{Hyperparameters}
\label{app:hyp}
Table~\ref{tab:train_hparams_combined} shows the hyperparameters used for training the encoders ST-MEM \citep{na2024guidingmaskedrepresentationlearning} and MERL \citep{liu2024zeroshotecgclassificationmultimodal}, as well as the ELMs for ELF, ST-MEM, and MERL.

\begin{table}
\centering
\small
\caption{Training hyperparameters used in this study. Left: ELF, ST-MEM, and MERL ELM training. Right: pretraining ST-MEM \citep{na2024guidingmaskedrepresentationlearning} and MERL \citep{liu2024zeroshotecgclassificationmultimodal} encoders. All trainings were conducted on 4$\times$H100 NVL NVIDIA GPUs.}
\label{tab:train_hparams_combined}

\begin{minipage}{0.47\textwidth}
\centering
\textbf{ELM Training Hyperparameters}

\vspace{0.4em}
\begin{tabular}{ll}
\toprule
\textbf{Hyperparameter} & \textbf{Value} \\
\midrule
Optimizer & Muon \\
Learning rate & $1 \times 10^{-3}$ \\
LR schedule & Cosine \\
Minimum LR ratio & 0.1 \\
Weight decay & 0.05 \\
$\beta_1$ & 0.9 \\
$\beta_2$ & 0.95 \\
$\epsilon$ & $1 \times 10^{-8}$ \\
Gradient clipping & 1.0 \\
Batch size & 4 \\
Reference global batch size & 16 \\
Gradient accumulation steps & 1 \\
Epochs & 10 \\
Patience & 5 \\
Patience delta & 0.1 \\
LoRA rank & 16 \\
LoRA alpha & 32 \\
LoRA dropout & 0.05 \\
LLM input length & 2048 \\
\bottomrule
\end{tabular}
\end{minipage}
\hfill
\begin{minipage}{0.47\textwidth}
\centering
\textbf{Encoder Pretraining Hyperparameters}

\vspace{0.4em}
\begin{tabular}{ll}
\toprule
\textbf{Hyperparameter} & \textbf{Value} \\
\midrule
Optimizer & AdamW \\
Learning rate & $3 \times 10^{-4}$ \\
LR schedule & Cosine \\
Minimum LR ratio & 0.1 \\
Weight decay & 0.01 \\
$\beta_1$ & 0.9 \\
$\beta_2$ & 0.95 \\
$\epsilon$ & $1 \times 10^{-8}$ \\
Gradient clipping & 1.0 \\
Batch size & 64 \\
Reference global batch size & 256 \\
Gradient accumulation steps & 1 \\
Epochs & 100 \\
Patience & 5 \\
Patience delta & 0.1 \\
\bottomrule
\end{tabular}
\end{minipage}

\end{table}

\begin{table}
\centering
\caption{Training time breakdowns for ELF, ST-MEM, and MERL. The MIMIC-IV-ECG dataset was used during the encoder training stage, while the ECG-Instruct 45K dataset was used for the ELM training stage.}
\label{tab:training_time}
\begin{tabular}{lcc}
\toprule
Model & Stage & Training Time \\
\midrule
\multirow{2}{*}{ST-MEM} & Encoder & 6h 45m \\
 & ELM & 3h 59m 21s \\
\midrule
\multirow{2}{*}{MERL} & Encoder & 8h 35m \\
 & ELM & 3h 43m 29s \\
\midrule
Base ELF & ELM & 2h 10m 40s \\
Patch ELF & ELM & 1h 50m 20s \\
Conv. ELF & ELM & 3h 14m 28s \\
\bottomrule
\end{tabular}
\end{table}
\begin{table}
\centering
\small
\caption{Comparing performance on ECG-QA-CoT between various baselines and ELF.}
\label{tab:ecgqa_cot_opentslm}
\resizebox{1.0\linewidth}{!}{%
\begin{tabular}{llccc ccc}
\toprule
ELM & LLM Backbone & BLEU-4 & ROUGE-L & METEOR & BERTScore F1 & F1 & Accuracy \\
\midrule
\multirow{4}{*}{OpenTSLM SoftPrompt \cite{langer2025opentslmtimeserieslanguagemodels}}
& Llama3.2-1B        & - & - & - & - & 32.84 & 35.49 \\
& Llama3.2-3B        & - & - & - & - & 33.67 & 36.25 \\
& Gemma3-270M        & - & - & - & - & 1.29  & 1.11  \\
& Gemma3-1B-pt       & - & - & - & - & 27.86 & 34.76 \\
\midrule
\multirow{4}{*}{OpenTSLM Flamingo \cite{langer2025opentslmtimeserieslanguagemodels}}
& Llama3.2-1B        & - & - & - & - & 34.62 & 38.14 \\
& Llama3.2-3B        & - & - & - & - & 40.25 & 46.25 \\
& Gemma3-270M        & - & - & - & - & 32.71 & 35.50 \\
& Gemma3-1B-pt       & - & - & - & - & 35.31 & 37.79 \\
\midrule
SLIP \cite{chen2026learningtransferablesensormodels} & Gemma3-270M & - & - & - & - & - & 37.18 \\
\midrule

\multirow{3}{*}{Base ELF}
& Llama-3.2-1B-Instruct   & 12.74 $\pm$ 0.06 & 47.06 $\pm$ 0.13 & 29.75 $\pm$ 0.12 & 91.94 $\pm$ 0.03 & 46.79 $\pm$ 0.13 & 42.46 $\pm$ 0.16 \\
& gemma-2-2b-it & 11.85 $\pm$ 0.02 & 46.58 $\pm$ 0.11 & 29.30 $\pm$ 0.09 & 91.90 $\pm$ 0.02 & 46.36 $\pm$ 0.10 & 42.25 $\pm$ 0.13 \\
& Qwen2.5-1.5B-Instruct   & 15.75 $\pm$ 0.02 & 39.92 $\pm$ 0.02 & 29.00 $\pm$ 0.04 & 90.42 $\pm$ 0.01 & 39.87 $\pm$ 0.01 & 33.36 $\pm$ 0.08 \\
\midrule

\multirow{3}{*}{Patch ELF}
& Llama-3.2-1B-Instruct   & 27.49 $\pm$ 0.12 & 59.21 $\pm$ 0.10 & 40.93 $\pm$ 0.11 & 93.93 $\pm$ 0.01 & 58.78 $\pm$ 0.10 & 54.63 $\pm$ 0.16 \\
& gemma-2-2b-it & \textbf{28.52 $\pm$ 0.08} & \textbf{62.17 $\pm$ 0.04} & \textbf{42.83 $\pm$ 0.07} & \textbf{94.34 $\pm$ 0.01} & \textbf{61.83 $\pm$ 0.05} & \textbf{58.55 $\pm$ 0.11} \\
& Qwen2.5-1.5B-Instruct   & 27.07 $\pm$ 0.00 & 62.10 $\pm$ 0.01 & 42.65 $\pm$ 0.01 & 94.31 $\pm$ 0.00 & 61.75 $\pm$ 0.01 & 58.36 $\pm$ 0.02 \\
\midrule

\multirow{3}{*}{Conv. ELF}
& Llama-3.2-1B-Instruct   & 27.95 $\pm$ 0.12 & 59.85 $\pm$ 0.04 & 41.50 $\pm$ 0.07 & 94.02 $\pm$ 0.01 & 59.41 $\pm$ 0.03 & 55.18 $\pm$ 0.02 \\
& gemma-2-2b-it & 26.06 $\pm$ 0.10 & 59.56 $\pm$ 0.03 & 40.92 $\pm$ 0.05 & 93.94 $\pm$ 0.01 & 59.22 $\pm$ 0.02 & 55.55 $\pm$ 0.01 \\
& Qwen2.5-1.5B-Instruct   & 24.75 $\pm$ 0.03 & 60.28 $\pm$ 0.00 & 41.16 $\pm$ 0.01 & 94.00 $\pm$ 0.01 & 59.93 $\pm$ 0.00 & 56.31 $\pm$ 0.00 \\

\bottomrule
\end{tabular}}
\end{table}

\begin{table}
\centering
\small
\caption{Comparing performance of ELF on ECG-Instruct 45K against ELMs with strong encoders.}
\label{tab:ecg_instruct}
\resizebox{1.0\linewidth}{!}{%
\begin{tabular}{llccc ccc}
\toprule
ELM & LLM Backbone & BLEU-4 & ROUGE-L & METEOR & BERTScore F1 & F1 & Accuracy \\
\midrule

MERL \cite{liu2024zeroshotecgclassificationmultimodal} & Llama-3.2-1B-Instruct & \textbf{39.17 $\pm$ 0.03} & \textbf{75.85 $\pm$ 0.02} & \textbf{73.16 $\pm$ 0.03} & \textbf{96.91 $\pm$ 0.00} & \textbf{78.06 $\pm$ 0.03} & \textbf{26.53 $\pm$ 0.09} \\

ST-MEM \cite{na2024guidingmaskedrepresentationlearning} & Llama-3.2-1B-Instruct & 37.02 $\pm$ 0.01 & 74.92 $\pm$ 0.02 & 72.09 $\pm$ 0.03 & 96.74 $\pm$ 0.00 & 77.13 $\pm$ 0.02 & 25.91 $\pm$ 0.09 \\

\midrule
Base ELF & Llama-3.2-1B-Instruct & 30.14 $\pm$ 0.07 & 70.97 $\pm$ 0.06 & 67.69 $\pm$ 0.08 & 96.09 $\pm$ 0.01 & 73.18 $\pm$ 0.06 & 20.80 $\pm$ 0.10 \\
Patch ELF & Llama-3.2-1B-Instruct & 30.50 $\pm$ 0.07 & 71.14 $\pm$ 0.08 & 67.72 $\pm$ 0.07 & 96.16 $\pm$ 0.01 & 73.33 $\pm$ 0.06 & 19.94 $\pm$ 0.01 \\
Conv. ELF & Llama-3.2-1B-Instruct & 33.60 $\pm$ 0.01 & 73.18 $\pm$ 0.01 & 70.20 $\pm$ 0.00 & 96.43 $\pm$ 0.00 & 75.40 $\pm$ 0.02 & 23.96 $\pm$ 0.07 \\

\bottomrule
\end{tabular}}
\end{table}

\begin{table}
\centering
\small
\caption{We evaluate Conv. ELF on the test set of ECG-QA-CoT on selected epochs.}
\label{tab:steps}
\resizebox{1.0\linewidth}{!}{%
\begin{tabular}{lcccccc}
\toprule
Epoch & BLEU-4 & ROUGE-L & METEOR & BERTScore F1 & F1 & Accuracy \\
\midrule
1 & 14.91 $\pm$ 0.20 & 43.25 $\pm$ 0.01 & 28.88 $\pm$ 0.01 & 91.46 $\pm$ 0.01 & 42.92 $\pm$ 0.02 & 37.16 $\pm$ 0.00 \\
2 & 20.42 $\pm$ 0.21 & 48.98 $\pm$ 0.05 & 33.62 $\pm$ 0.07 & 92.59 $\pm$ 0.02 & 48.28 $\pm$ 0.08 & 42.35 $\pm$ 0.03 \\
4 & 27.27 $\pm$ 0.05 & 55.91 $\pm$ 0.22 & 39.20 $\pm$ 0.20 & 93.56 $\pm$ 0.03 & 55.43 $\pm$ 0.23 & 50.05 $\pm$ 0.23 \\
6 & \textbf{29.67 $\pm$ 0.28} & 57.98 $\pm$ 0.40 & 41.08 $\pm$ 0.31 & 93.85 $\pm$ 0.05 & 57.59 $\pm$ 0.39 & 52.85 $\pm$ 0.45 \\
8 & 27.77 $\pm$ 0.05 & 59.13 $\pm$ 0.04 & 40.75 $\pm$ 0.01 & 93.94 $\pm$ 0.00 & 58.74 $\pm$ 0.04 & 54.69 $\pm$ 0.07 \\
10 & 27.95 $\pm$ 0.12 & \textbf{59.85 $\pm$ 0.04} & \textbf{41.50 $\pm$ 0.07} & \textbf{94.02 $\pm$ 0.01} & \textbf{59.41 $\pm$ 0.03} & \textbf{55.18 $\pm$ 0.02} \\
\bottomrule
\end{tabular}}
\end{table}
\begin{table}
\centering
\small
\caption{We evaluate Conv. ELF on the test set of ECG-Instruct 45K on selected epochs.}
\label{tab:steps2}
\resizebox{1.0\linewidth}{!}{%
\begin{tabular}{lcccccc}
\toprule
Epoch & BLEU-4 & ROUGE-L & METEOR & BERTScore F1 & F1 & Accuracy \\
\midrule
1  & 26.69 $\pm$ 0.01 & 68.11 $\pm$ 0.00 & 64.20 $\pm$ 0.01 & 95.71 $\pm$ 0.00 & 70.35 $\pm$ 0.01 & 14.97 $\pm$ 0.00 \\
2  & 29.26 $\pm$ 0.03 & 70.14 $\pm$ 0.01 & 66.53 $\pm$ 0.01 & 95.98 $\pm$ 0.00 & 72.37 $\pm$ 0.00 & 17.98 $\pm$ 0.18 \\
4  & 32.94 $\pm$ 0.11 & 72.41 $\pm$ 0.06 & 69.37 $\pm$ 0.08 & 96.38 $\pm$ 0.01 & 74.68 $\pm$ 0.07 & 21.75 $\pm$ 0.00 \\
6  & 33.57 $\pm$ 0.03 & 73.11 $\pm$ 0.00 & 70.15 $\pm$ 0.03 & 96.44 $\pm$ 0.00 & 75.35 $\pm$ 0.01 & 23.78 $\pm$ 0.06 \\
8  & \textbf{33.83 $\pm$ 0.02} & \textbf{73.27 $\pm$ 0.00} & \textbf{70.37 $\pm$ 0.03} & \textbf{96.45 $\pm$ 0.01} & \textbf{75.51 $\pm$ 0.02} & \textbf{24.08 $\pm$ 0.14} \\
10 & 33.60 $\pm$ 0.01 & 73.18 $\pm$ 0.01 & 70.20 $\pm$ 0.00 & 96.43 $\pm$ 0.00 & 75.40 $\pm$ 0.02 & 23.96 $\pm$ 0.07 \\
\bottomrule
\end{tabular}}
\end{table}
\begin{table}
\centering
\caption{We selectively train Patch ELF's components and report mean $\pm$ standard deviation over three seeds on ECG-QA-CoT across two LLM backbones. \checkmark indicates the corresponding component was set as trainable.}
\label{tab:freeze_ecgqa}
{\resizebox{1\textwidth}{!}{%
\begin{tabular}{lccccccc}
\toprule
\multirow{2}{*}{LLM Backbone} &
\multicolumn{2}{c}{Components} &
\multirow{2}{*}{BLEU-4} &
\multirow{2}{*}{ROUGE-L} &
\multirow{2}{*}{METEOR} &
\multirow{2}{*}{F1} &
\multirow{2}{*}{Accuracy} \\
\cmidrule(r){2-3}
& LLM & Projection Layer $W$ & & & & & \\
\midrule

\multirow{3}{*}{Llama-3.2-1B-Instruct}
& \checkmark &  & 26.81 $\pm$ 0.05 & \textbf{61.83 $\pm$ 0.01} & \textbf{42.47 $\pm$ 0.03} & \textbf{61.49 $\pm$ 0.01} & \textbf{57.97 $\pm$ 0.02} \\
&  & \checkmark & 13.94 $\pm$ 0.05 & 43.99 $\pm$ 0.43 & 28.77 $\pm$ 0.22 & 43.54 $\pm$ 0.43 & 38.20 $\pm$ 0.41 \\
& \checkmark & \checkmark & \textbf{27.49 $\pm$ 0.12} & 59.21 $\pm$ 0.10 & 40.93 $\pm$ 0.11 & 58.78 $\pm$ 0.10 & 54.63 $\pm$ 0.16  \\
\midrule

\multirow{3}{*}{gemma-2-2b-it}
& \checkmark &  & \textbf{30.10 $\pm$ 0.03} & \textbf{62.98 $\pm$ 0.02} & \textbf{43.81 $\pm$ 0.05} & \textbf{62.64 $\pm$ 0.10} & \textbf{59.15 $\pm$ 0.07} \\
&  & \checkmark & 18.31 $\pm$ 0.06 & 49.00 $\pm$ 0.03 & 33.01 $\pm$ 0.08 & 48.60 $\pm$ 0.05 & 42.36 $\pm$ 0.05 \\
& \checkmark & \checkmark & 28.52 $\pm$ 0.08 & 62.17 $\pm$ 0.04 & 42.83 $\pm$ 0.07 & 61.83 $\pm$ 0.05 & 58.55 $\pm$ 0.11 \\

\bottomrule
\end{tabular}}}
\end{table}

\begin{table}
\centering
\caption{We selectively train Patch ELF's components and report mean $\pm$ standard deviation over three seeds on ECG-Instruct 45K using Llama-3.2-1B-Instruct. \checkmark indicates the corresponding component was set as trainable.}
\label{tab:freeze_ecginstruct}
{\resizebox{1\textwidth}{!}{%
\begin{tabular}{lccccccc}
\toprule
\multicolumn{2}{c}{Components} &
\multirow{2}{*}{BLEU-4} &
\multirow{2}{*}{ROUGE-L} &
\multirow{2}{*}{METEOR} &
\multirow{2}{*}{F1} &
\multirow{2}{*}{Accuracy} \\
\cmidrule(r){1-2}
LLM & Projection Layer $W$ & & & & & \\
\midrule


\checkmark &  & \textbf{31.76 $\pm$ 0.02} & \textbf{72.24 $\pm$ 0.01} & \textbf{68.85 $\pm$ 0.01} & \textbf{74.38 $\pm$ 0.01} & \textbf{22.58 $\pm$ 0.11} \\
& \checkmark & 19.53 $\pm$ 0.03 & 59.16 $\pm$ 0.05 & 54.95 $\pm$ 0.05 & 61.47 $\pm$ 0.04 & 6.41 $\pm$ 0.07 \\
\checkmark & \checkmark &30.50 $\pm$ 0.07 & 71.14 $\pm$ 0.08 & 67.72 $\pm$ 0.07  & 73.33 $\pm$ 0.06 & 19.94 $\pm$ 0.01 \\
\bottomrule
\end{tabular}}}
\end{table}

\begin{table}[t]
\centering
\small
\caption{ECG-QA-CoT results when varying the number of encoder tokens $N = \{50, 100\}$ for Patch ELF and Conv.\ ELF.}
\label{tab:patch_conv_elf}
\resizebox{1.0\linewidth}{!}{%
\begin{tabular}{llcccccc}
\toprule
Model & Tokens & BLEU-4 & ROUGE-L & METEOR & BERTScore & F1 &Accuracy \\
\midrule
\multirow{2}{*}{Patch ELF}
 & 50  & \textbf{27.49 $\pm$ 0.12} & \textbf{59.21 $\pm$ 0.10} & \textbf{40.93 $\pm$ 0.11} & \textbf{93.93 $\pm$ 0.01} & \textbf{58.78 $\pm$ 0.10} & \textbf{54.63 $\pm$ 0.16} \\
 & 100 & 25.88 $\pm$ 0.02 & 57.06 $\pm$ 0.07 & 39.04 $\pm$ 0.09 & 93.59 $\pm$ 0.01 & 56.66 $\pm$ 0.08 & 52.67 $\pm$ 0.09 \\
\midrule
\multirow{2}{*}{Conv.\ ELF}
 & 50  & \textbf{27.95 $\pm$ 0.12} & \textbf{59.85 $\pm$ 0.04} & \textbf{41.50 $\pm$ 0.07} & \textbf{94.02 $\pm$ 0.01} & \textbf{59.41 $\pm$ 0.03} & \textbf{55.18 $\pm$ 0.02} \\
 & 100 & 21.67 $\pm$ 0.07 & 56.93 $\pm$ 0.05 & 37.83 $\pm$ 0.04 & 93.47 $\pm$ 0.01 & 56.64 $\pm$ 0.06 & 52.79 $\pm$ 0.04 \\
\bottomrule
\end{tabular}}
\end{table}
\begin{table}
\centering
\small
\caption{Performance under different learning rates.}
\label{tab:lr}
\begin{tabular}{lccccc}
\toprule
Learning rate & BLEU-4 & ROUGE-L & METEOR & F1 & Accuracy \\
\midrule
1e-3 & \textbf{28.52 $\pm$ 0.08} & \textbf{62.17 $\pm$ 0.04} & \textbf{42.83 $\pm$ 0.07} & \textbf{61.83 $\pm$ 0.05} & \textbf{58.55 $\pm$ 0.11} \\
1e-4 & 21.02 $\pm$ 0.05 & 45.96 $\pm$ 0.04 & 33.70 $\pm$ 0.10 & 45.89 $\pm$ 0.06 & 40.32 $\pm$ 0.09 \\
1e-5 & 21.24 $\pm$ 0.07 & 46.84 $\pm$ 0.01 & 34.20 $\pm$ 0.09 & 46.42 $\pm$ 0.04 & 40.41 $\pm$ 0.10 \\
\bottomrule
\end{tabular}
\end{table}
\begin{table}
\centering
\small
\caption{ELF's performance under different input perturbation settings on the ECG-QA-CoT dataset}
\label{tab:perturbation_results}
\begin{tabular}{lccccc}
\toprule
Perturbation & BLEU-4 & ROUGE-L & METEOR & F1 & Accuracy \\
\midrule
Normal   & \textbf{28.52 $\pm$ 0.08} & \textbf{62.17 $\pm$ 0.04} & \textbf{42.83 $\pm$ 0.07} & \textbf{61.83 $\pm$ 0.05} & \textbf{58.55 $\pm$ 0.11} \\
Only Text & 8.20  $\pm$ 0.04 & 44.72 $\pm$ 0.04 & 27.28 $\pm$ 0.06 & 44.30 $\pm$ 0.05 & 39.94 $\pm$ 0.07 \\
Zeros     & 8.72  $\pm$ 0.02 & 44.76 $\pm$ 0.05 & 27.70 $\pm$ 0.05 & 44.33 $\pm$ 0.06 & 39.78 $\pm$ 0.06 \\
Noise     & 15.61 $\pm$ 0.07 & 43.55 $\pm$ 0.05 & 28.36 $\pm$ 0.07 & 43.36 $\pm$ 0.06 & 37.84 $\pm$ 0.08 \\
\bottomrule
\end{tabular}
\end{table}

\begin{figure}
    \centering
    \includegraphics[width=1\linewidth]{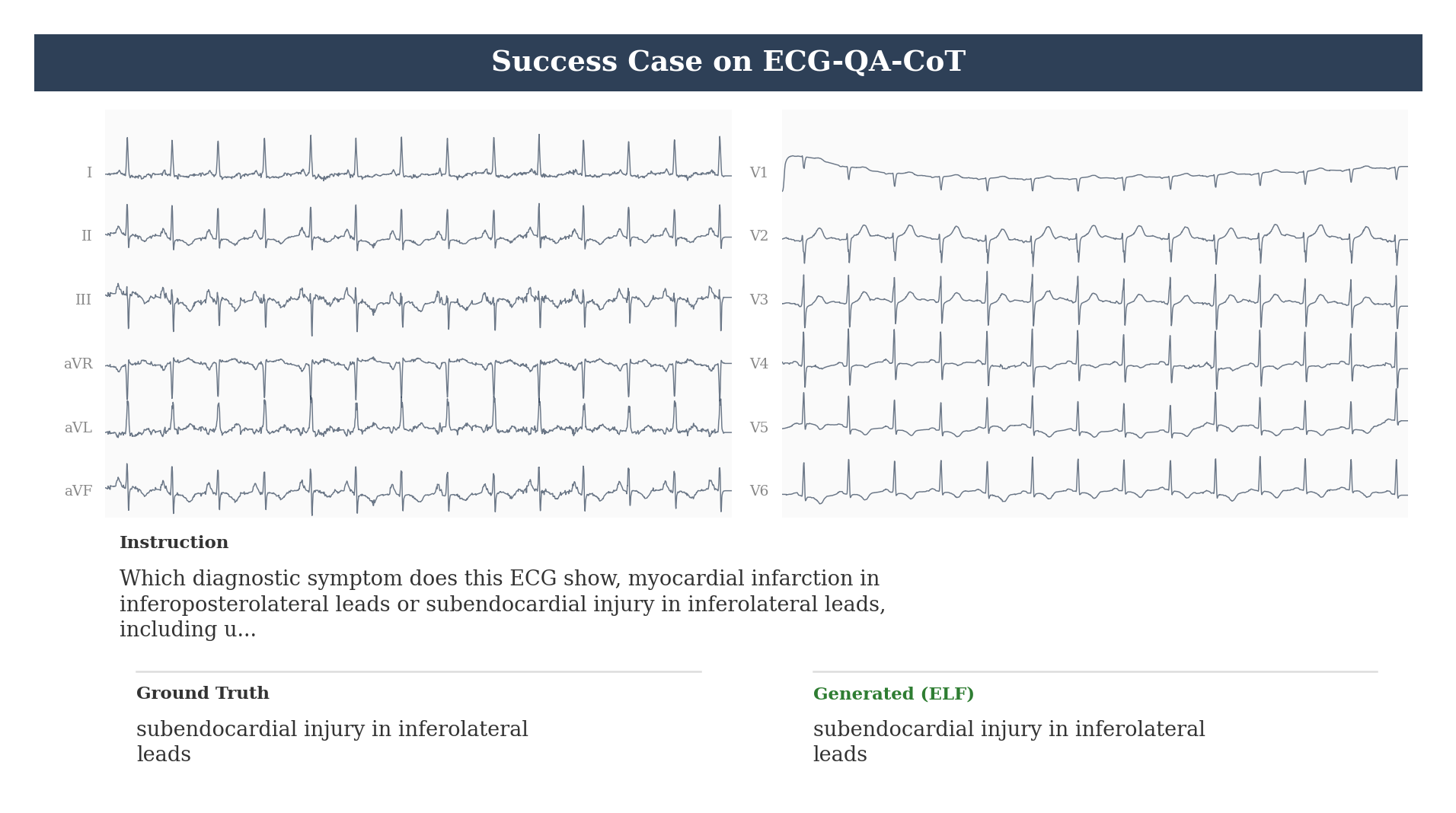}
    \caption{A successful generation with Conv. ELF on the ECG-QA-CoT dataset}
    \label{fig:ecg-qa-success}
\end{figure}

\begin{figure}
    \centering
    \includegraphics[width=1\linewidth]{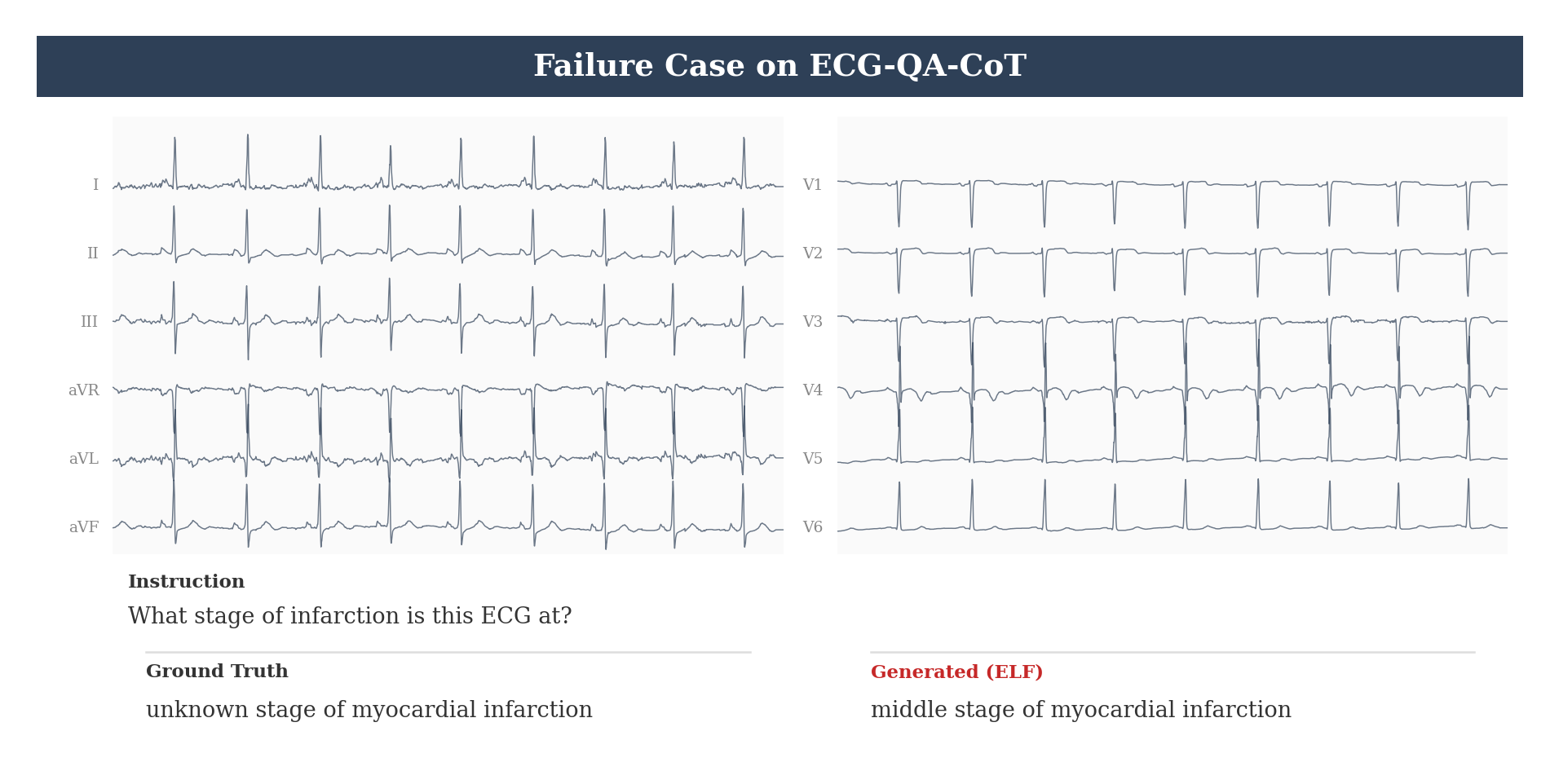}
    \caption{Failed generation with Conv. ELF on the ECG-QA-CoT dataset}
    \label{fig:ecg-qa-fail}
\end{figure}

\begin{figure}
    \centering
    \includegraphics[width=1\linewidth]{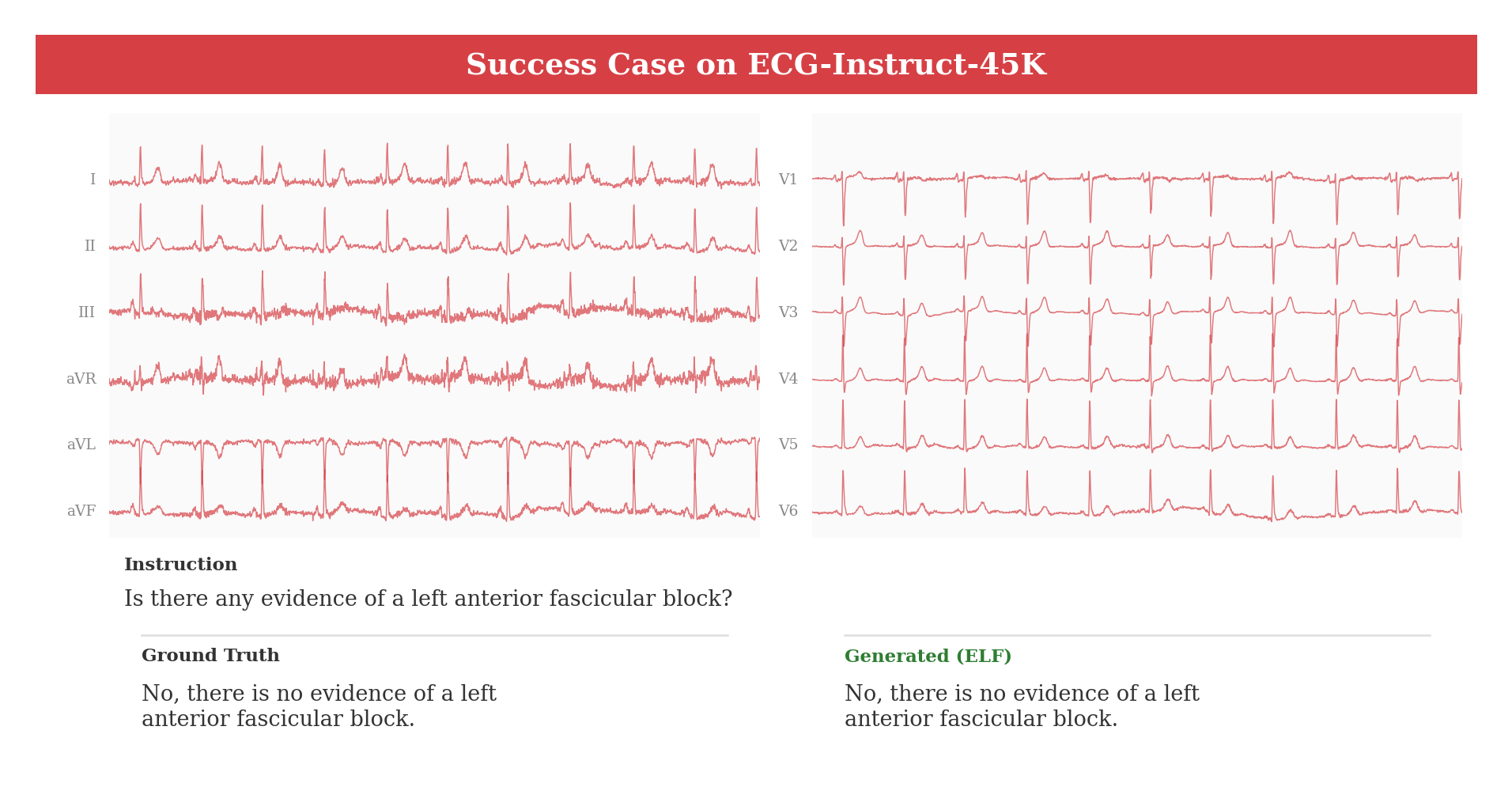}
    \caption{A successful generation with Conv. ELF on the ECG-Instruct 45K dataset}
    \label{fig:ecg-instruct-success}
\end{figure}

\begin{figure}
    \centering
    \includegraphics[width=1\linewidth]{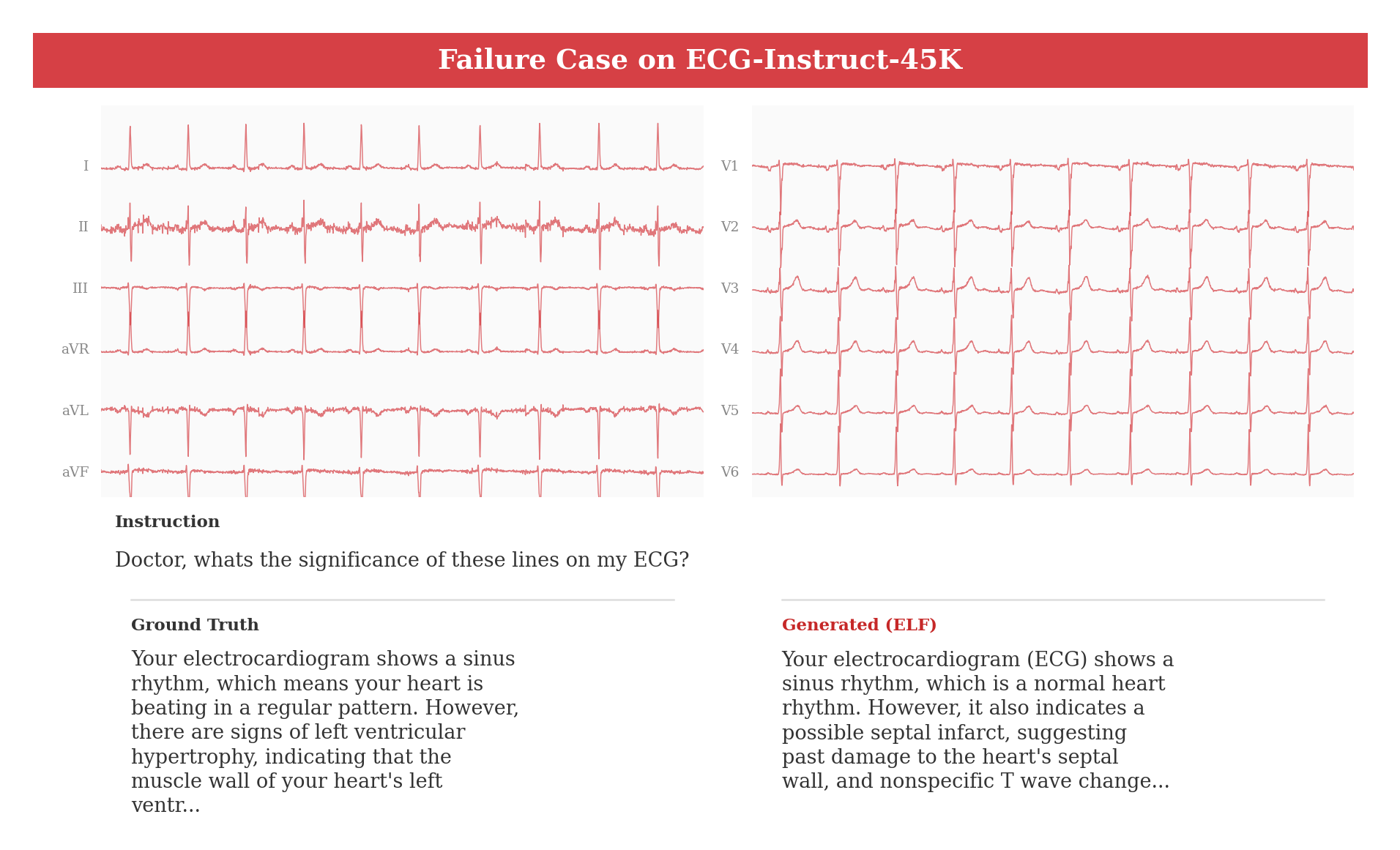}
    \caption{Failed generation with Conv. ELF on the ECG-Instruct 45K dataset}
    \label{fig:ecg-instruct-fail}
\end{figure}
\newpage
\section{Additional Results}
\label{app:add_results}
We provide the tabular form with additional metrics for each figure in the main paper. Specifically, Table~\ref{tab:ecgqa_cot_opentslm} corresponds to Figure~\ref{fig:main}, Table~\ref{tab:ecg_instruct} corresponds to Figure~\ref{fig:analysis2}, Tables~\ref{tab:steps} and \ref{tab:steps2} correspond to Figure~\ref{fig:analysis}, Tables~\ref{tab:freeze_ecgqa}, \ref{tab:freeze_ecginstruct}, and \ref{tab:patch_conv_elf} correspond to Figure~\ref{fig:analysis3}, and Table~\ref{tab:perturbation_results} corresponds to Figure~\ref{fig:perturb}.

\subsection{Learning Rate}
In Table~\ref{tab:lr}, we present additional experiments varying the learning rate when using Patch ELF, gemma-2-2b-it on the ECG-QA-CoT dataset. We observe our original learning rate (1e-3) consistently outperforms 1e-4 and 1e-5 across all metrics.

\subsection{Dataset Instance Visualizations}
\label{app:data_inst}
\begin{figure}
    \centering
    \includegraphics[width=0.75\linewidth]{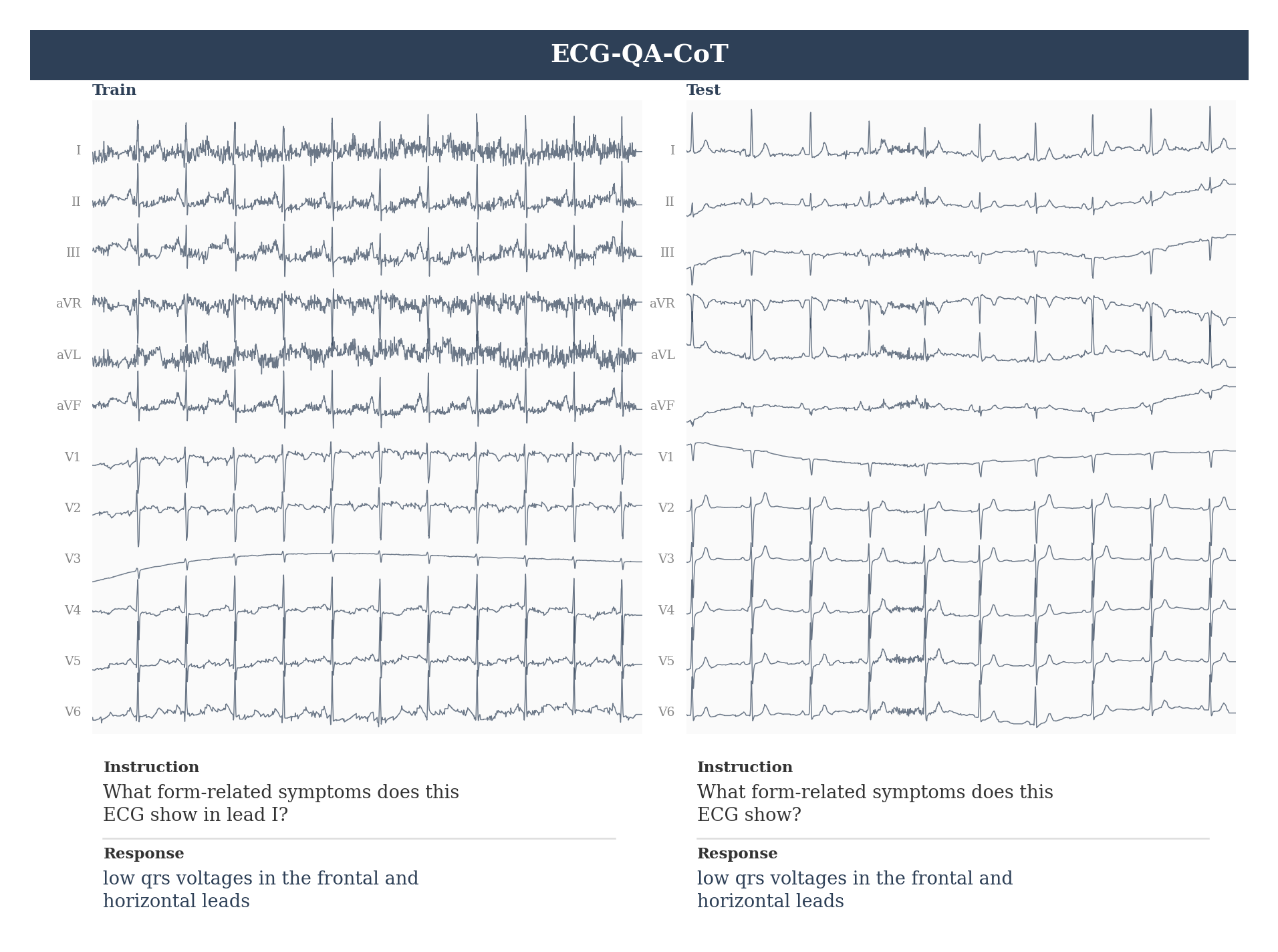}
    \caption{Instances from the training and testing set of ECG-QA-CoT.}
    \label{fig:ecg-qa-inst}
\end{figure}

\begin{figure}
    \centering
    \includegraphics[width=0.75\linewidth]{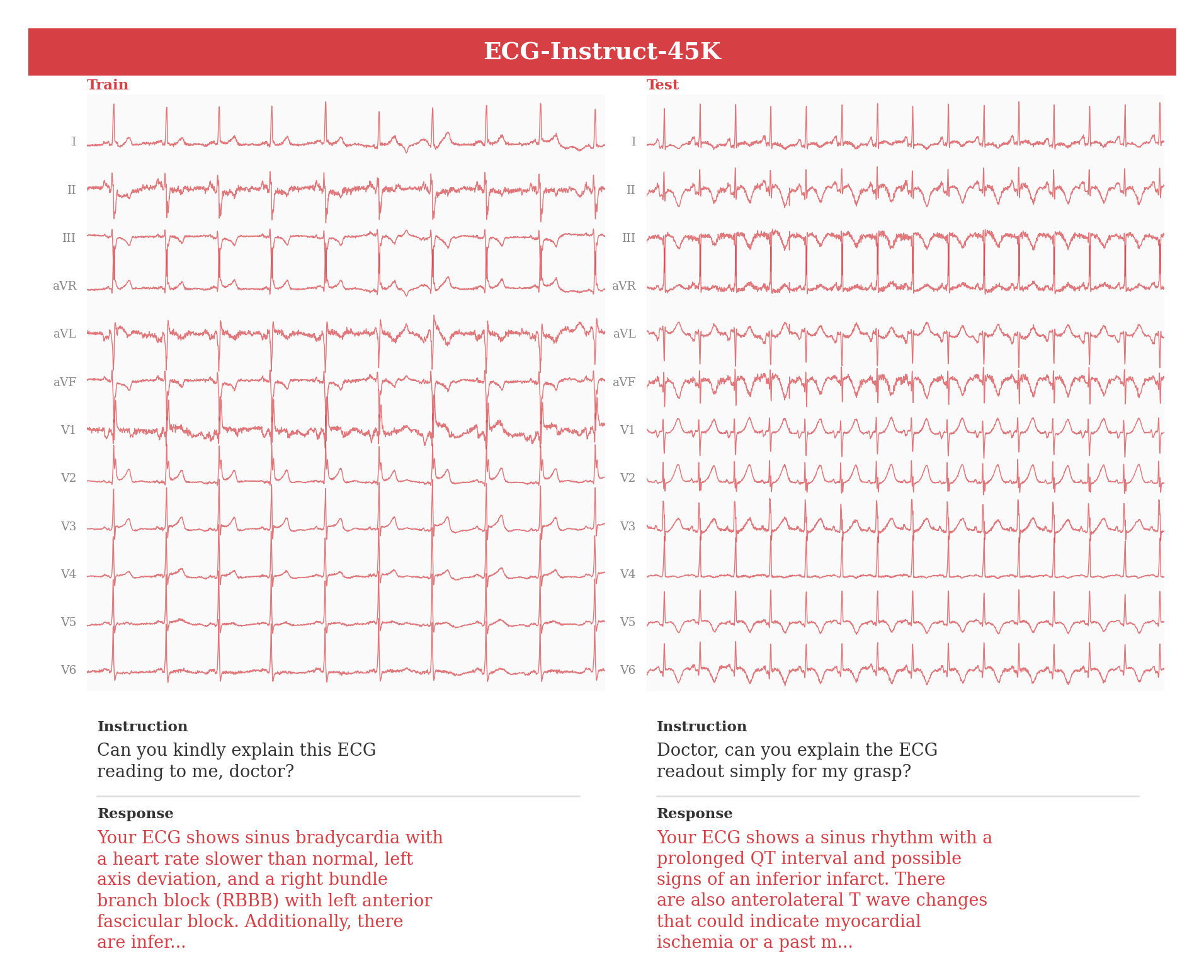}
    \caption{Instances from the training and testing set of ECG-Instruct 45K.}
    \label{fig:ecg-instruct-inst}
\end{figure}

We visualize examples of instances from the ECG-QA-CoT and ECG-Instruct 45K datasets in Figure~\ref{fig:ecg-qa-inst} and \ref{fig:ecg-instruct-inst} respectively.

\section{Templates}
\label{sec:conv_temp}
\subsection{Large Language Model Conversation Templates}
We provide the conversation templates for Llama-3.2-1B-Instruct, gemma-2-2b-it, and Qwen-2.5-1.5B-Instruct. $q_{\text{sys}}$, $q_n$, $y_n$ refers to the system prompt, textual query and response respectively.

\begin{tcolorbox}[colback=white, colframe=black, sharp corners, boxrule=0.5mm]
\textbf{Llama-3.2-1B-Instruct Conversation Template}\\\\
$<|\text{begin\_of\_text}|><|\text{start\_header\_id}|>\text{system}<|\text{end\_header\_id}|>$\\\\
$q_{\text{sys}}<|\text{eot\_id}|><|\text{start\_header\_id}|>\text{user}<|\text{end\_header\_id}|>$\\\\
\texttt{<signal>}\\
$q_1<|\text{eot\_id}|><|\text{start\_header\_id}|>\text{assistant}<|\text{end\_header\_id}|>$\\\\

$y_1<|\text{eot\_id}|><|\text{start\_header\_id}|>\text{user}<|\text{end\_header\_id}|>$\\\\

$q_2<|\text{eot\_id}|><|\text{start\_header\_id}|>\text{assistant}<|\text{end\_header\_id}|>$\\\\

$y_2<|\text{eot\_id}|>$

... \\

$q_n<|\text{eot\_id}|><|\text{start\_header\_id}|>\text{assistant}<|\text{end\_header\_id}|>$\\\\

$y_n<|\text{eot\_id}|>$
\end{tcolorbox}
\newpage
\begin{tcolorbox}[colback=white, colframe=black, sharp corners, boxrule=0.5mm]
\textbf{gemma-2-2b-it Conversation Template}\\\\
$<\text{bos}><\text{start\_of\_turn}>\text{user}$\\
\texttt{<signal>}\\
$q_1<\text{end\_of\_turn}>$\\
$<\text{start\_of\_turn}>\text{model}$\\
$y_1<\text{end\_of\_turn}>$\\
$<\text{start\_of\_turn}>\text{user}$\\
$q_2<\text{end\_of\_turn}>$\\
$<\text{start\_of\_turn}>\text{model}$\\
$y_2<\text{end\_of\_turn}>$\\

... \\

$<\text{start\_of\_turn}>\text{user}$\\
$q_n<\text{end\_of\_turn}>$\\
$<\text{start\_of\_turn}>\text{model}$\\
$y_n<\text{end\_of\_turn}>$\\
\end{tcolorbox}

\begin{tcolorbox}[colback=white, colframe=black, sharp corners, boxrule=0.5mm]
\textbf{Qwen-2.5-1.5B-Instruct Conversation Template}\\\\
$<|\text{im\_start}|>\text{system}$\\
$q_{\text{sys}}<|\text{im\_end}|>$\\
$<|\text{im\_start}|>\text{user}$\\
\texttt{<signal>}\\
$q_1<|\text{im\_end}|>$\\
$<|\text{im\_start}|>\text{assistant}$\\
$y_1<|\text{im\_end}|>$\\
$<|\text{im\_start}|>\text{user}$\\
$q_2<|\text{im\_end}|>$\\
$<|\text{im\_start}|>\text{assistant}$\\
$y_2<|\text{im\_end}|>$\\

... \\

$<|\text{im\_start}|>\text{user}$\\
$q_n<|\text{im\_end}|>$\\
$<|\text{im\_start}|>\text{assistant}$\\
$y_n<|\text{im\_end}|>$\\
\end{tcolorbox}

\newpage
\subsection{System Prompt}
\label{sec:sys_prompt}
We provide the system prompts utilized throughout the study. Namely, we provide the ELM system prompt used for all experiments with ELF, ST-MEM \cite{na2024guidingmaskedrepresentationlearning}, and MERL \cite{liu2024zeroshotecgclassificationmultimodal}.

\begin{tcolorbox}[colback=white, colframe=black, sharp corners, boxrule=0.5mm]
\textbf{ELM System Prompt}\\
You are an expert multimodal assistant with advanced knowledge in **clinical cardiac electrophysiology**.\\

**Input Identification**\\
- Detect whether the input is **text**, **ECG signals (time-series data)**, or **both**.\\

**ECG Signal Analysis**\\
- Treat ECG as cardiac time-series data.\\
- Provide expert interpretation of **heart rate, rhythm, conduction, arrhythmias, and other electrophysiologic abnormalities**.\\

**Multimodal Reasoning**\\
- When both text and ECG are given, integrate them into a unified, **cardiac electrophysiologist-level assessment**.\\

**Response Style**\\
- Deliver responses that are **precise, concise, and clinically authoritative**, grounded in electrophysiology and natural language reasoning.\\
- For general, non-ECG questions, respond as a capable **general assistant**.
\end{tcolorbox}

\end{document}